\documentclass[superscriptaddress,groupedaddress,nofootnoteinbib,11pt]{article}
\pdfoutput=1 
\usepackage{jheppub}

\usepackage{amsmath, amsfonts, amsthm, amssymb, graphicx, color, hyperref, slashed}
\usepackage{subfigure}
\usepackage{pstool}
\usepackage{graphicx}
\usepackage{float}
\usepackage[utf8]{inputenc}
\usepackage[normalem]{ulem}

\allowdisplaybreaks[3]

\def \bal#1\eal  {\begin{align} #1 \end{align}}
\def\({\left(}
\def\){\right)}
\def\[{\left[}
\def\]{\right]}
\def\<{\left\langle}
\def\>{\right\rangle}
\def\d{\mathrm{d}}

\newcommand{\eref}[1]{Eq.~(\ref{#1})}
\newcommand{\f}[2]{\frac{#1}{#2}}
\newcommand{\bim} {\begin{itemize}}

\newcommand{\eim}{\end{itemize}}
\newcommand{\be} {\begin{equation}}
\newcommand{\ee} {\end{equation}}
\newcommand{\bc}{\begin{center}}
\newcommand{\ec}{\end{center}}

\newcommand{\nn} {\nonumber\\}

\newcommand{\ddt}{\frac{\d}{\d t}}

\newcommand{\dd} {\delta}

\newcommand{\pd} {\partial}
\newcommand{\mc} {\mathcal}

\newcommand{\ai}{{\alpha}}
\newcommand{\bi}{{\beta}}
\newcommand{\gi}{{\gamma}}
\newcommand{\di}{{\delta}}
\newcommand{\ri}{{\rho}}
\newcommand{\si}{{\sigma}}
\newcommand{\li}{{\lambda}}

\newcommand{\ei}{{\eta}}

\newcommand{\epi}{\epsilon}
\newcommand{\thi}{\theta}
\newcommand{\Gi}{\Gamma}
\newcommand{\Li}{\Lambda}
\def \= {\equiv}
\newcommand{\beq} {\begin{equation}}
\newcommand{\eeq} {\end{equation}}

\newcommand{\ki}{\kappa}

\title{Generalized elastic positivity bounds on interacting massive spin-2 theories}

\author[a]{Zi-Yue Wang,}
\author[b,c]{Cen Zhang,}
\author[d,e]{and Shuang-Yong Zhou}
\affiliation[a]{School of Gifted Young, University of Science and Technology of China, Hefei, Anhui 230026, China}
\affiliation[b]{
Institute for High Energy Physics, and School of Physical Sciences, University
of Chinese Academy of Sciences, Beijing 100049, China
}
\affiliation[c]{Center for High Energy Physics, Peking University, Beijing 100871, China}
\affiliation[d]{Interdisciplinary Center for Theoretical Study, University of Science and Technology of China, Hefei, Anhui 230026, China}
\affiliation[e]{Peng Huanwu Center for Fundamental Theory, Hefei, Anhui 230026, China}
\emailAdd{metrictensor@mail.ustc.edu.cn}
\emailAdd{cenzhang@ihep.ac.cn}
\emailAdd{zhoushy@ustc.edu.cn}

\preprint{\footnotesize USTC-ICTS/PCFT-20-39}

\date{\today}

\abstract{We use generalized elastic positivity bounds to constrain the parameter space of multi-field spin-2 effective field theories. These generalized bounds involve inelastic scattering amplitudes between particles with different masses, which contain kinematic singularities even in the $t=0$ limit. We apply these bounds to the pseudo-linear spin-2 theory, the cycle spin-2 theory and the line spin-2 theory respectively. For the pseudo-linear theory, we exclude the remaining operators that are unconstrained by the usual elastic positivity bounds, thus excluding all the leading (or highest cutoff) interacting operators in the theory. For the cycle and line theory, our approach also provides new bounds on the Wilson coefficients previously unconstrained, bounding the parameter space in both theories to be a finite region ({\it i.e.}, every Wilson coefficient being constrained from both sides). To help visualize these finite regions, we sample various cross sections of them and estimate the total volumes.}

\begin{document}
\maketitle
\flushbottom

\section{Introduction and summary}

General relativity may be viewed as a spin-2 effective field theory (EFT) around Minkowski space for a massless graviton. While the theory is very successful in describing gravitational interactions in the solar system and beyond, it is much less tested at large distances. Indeed, current cosmological observations suggest that our universe is accelerating, which might signal that gravity becomes weaker at cosmological scales, for example, due to the graviton actually having a small mass; See \cite{deRham:2016nuf} for a review of the graviton mass bounds from various observations. Massive spin-2 fields that couple to a massless graviton can also be dark matter candidates \cite{Babichev:2016bxi, Marzola:2017lbt, Bernal:2018qlk}, accounting for the deviations of galaxy rotation curves from general relativity. When higher dimensional gravity is compactified to a lower dimensional one via the Kaluza-Klein reduction, we get a tower of coupled massive spin-2 fields. Massive spin-2 fields are also relevant in non-gravity related contexts; For example, they appear as bound states in hadron physics and in condensed matter physics \cite{Gromov:2017qeb}.

There has been a long history of constructing massive spin-2 theories or massive gravity models. Fierz and Pauli wrote down the linear ghost-free massive spin-2 theory as early as 1930s \cite{Fierz:1939ix}. Boulware and Deser found that generic nonlinear massive gravity models contain a ghost degree of freedom (the BD ghost) \cite{Boulware:1973my}. The ghost-free version of massive gravity was discovered recently, known the dRGT model \cite{deRham:2010kj}, and the bi-gravity and multi-gravity extensions of the dRGT model have been formulated in \cite{Hassan:2011zd, Hinterbichler:2012cn}. See \cite{deRham:2014zqa, Hinterbichler:2011tt, Schmidt-May:2015vnx} for a review of the recent developments in massive gravity and multi-gravity theories. 
From the EFT point of view, having a ghost degree of freedom {\it per se} is not a cause for concern as long as the mass scale of the ghost is at or above the cutoff of the EFT. However, generic massive spin-2 theories typically have a very low EFT cutoff, which is accompanied by a range of problems \cite{Hinterbichler:2011tt}. In this language, the ghost-free dRGT model and its multi-field generalizations are simply the leading terms in a massive spin-2 EFT with the highest cutoff\;\footnote{The cutoff may be further raised if one considers a nearly flat but non-Minkowski background \cite{deRham:2016plk} or the theory is embedded in a braneworld setting \cite{Gabadadze:2017jom}.}, on the backdrop of an infinite tower of subleading higher derivative terms \cite{Alberte:2019lnd}. According to different topologies in the field space of spin-2 fields, generic multi-field spin-2 EFTs with the highest cutoff can be divided into two categories: the cycle theory and the line theory. For the cycle theory, different spin-2 fields are freely coupled to each other, and cycles are formed in the field space and/or non-pairwise interactions are present. The line theory, on the other hand, does not contain cycles or non-pairwise interactions, and thus only allows interactions between spin-2 fields neighboring in the field space. This is easiest to see in the vierbtein formulation of multi-gravity \cite{Hinterbichler:2012cn}, in which a generic multi-field massive spin-2 action in 4D with the highest cutoff is given by
\bal
\label{Smspin2more}
g_{*}^{2} S_{\rm ms}&=
\sum_a\frac{M_{a}^{2}}{8} \int \!\epsilon_{ABCD} H_{(a)}^{A} \wedge H_{(a)}^{B} \wedge R^{CD}[H_{(a)}]
 +\!\!\! \sum_{a,b,c,d=0}^N\!\!\! T^{abcd}\epi_{ABCD} H_{(a)}^A\wedge H_{(b)}^B\wedge H_{(c)}^C\wedge H_{(d)}^D +...  ,
\eal
where $\epsilon_{ABCD}$ is the flat space Levi-Civita tensor, $R^{AB}$ is the curvature two form,  $T^{abcd}$ is a constant tensor symmetric in its indices, $H_{(a)}^A$ are the vierbein fields, $M_a$ (all assumed to be around $M$) are the strong coupling scales of the helicity-2 modes of the corresponding vierbeins, $g_*$ is a dimensionless weak coupling, and $...$ stands for the higher derivative terms. Note that now the background Minkowski space also counts as a spin-2 field (albeit a trivial one) whose vierbein is the identity metric. The usual Planck scale $M_P$ is related to $M$ via $M_{\mathrm{P}}=M / g_{*}$. While the first part is a sum of the Einstein-Hilbert terms for different spin-2 fields, the second part is the dRGT potential for multiple vierbeins (the $a=b=c=d$ term being simply the cosmological constant), sharing the same double Levi-Civita structure as the Lovelock terms (here the Einstein-Hilbert term in 4D). In this formulation, the cycle theory is the generic theory where $T^{abcd}$ can be any constant tensor, while the line theory is when $T^{abcd}$ is chosen such that for any two spin-2 fields there is only one interaction term between term in the $abcd$ summation above.

To pass to the metric formulation, in the cycle theory, we will supply action (\ref{Smspin2more}) with symmetric vierbein conditions
\be
\label{cyclevierb}
\eta_{A[B}H_{(a)}{}^A{}_{\mu]} = 0 ,
\ee
where we have chosen $H_{(0)}^{A}$ to be the flat space vierbein, the identity matrix, so that we can expand all other metrics around Minkowski space. Using the vierbein perturbations $g^{(a)}_{\mu\nu} = (\eta_{\mu\ri}+{h^{(a)}{}_{\mu\ri}}/{M_a})\eta^{\ri\si} (\eta_{\si\nu}+{h^{(a)}{}_{\si\nu}}/{M_a})$,
the cycle theory for the case with two dynamical fields can be written as \eref{cycle}. The reason why the vierbein perturbations rather than the standard metric perturbations are used is to simplify the square root construction of the dRGT potential. This does mean that the Einstein-Hilbert term will be more complicated around Minkowski space, but this is a lesser price to pay for our purposes --- to compute tree level amplitudes for 2-to-2 scatterings. The cutoff of a generic cycle theory is at $\Lambda_{7/2}=(m^{5/2}M)^{2/7}$, with $m$ the mass scale of spin-2 particles, assuming that quadratic mass mixings are absent and cubic interactions are non-vanishing. If the vertices between different spin-2 fields are suppressed by an extra factor of $m/\Lambda_3$, with $\Lambda_3=(m^2M)^{1/3}$, then the cutoff is raised to $\Lambda_3$. The cycle theory is not free of the BD ghost \cite{Hinterbichler:2012cn, Scargill:2014wya}, which however is harmless as its mass is heavy \cite{Alberte:2019lnd}. (One may also consider the vierbein cycle theory without the symmetric vierbein conditions, which is {\it inequivalent} to our the metric cycle theory we consider in this paper and which still has the BD ghost \cite{deRham:2015cha}.) The line theory, however, can pass to the metric theory without imposing the symmetric vierbein conditions. Its symmetric vierbein conditions, which are different from those of \eref{cyclevierb}, can be obtained by integrating out the local Lorentz degrees of freedom, like that in the dRGT model or general relativity.  Still using the vierbein perturbations as in the cycle theory, the line theory can be written as \eref{linehf} in terms of the mass eigenmodes. The EFT cutoff of the line theory is at $\Lambda_3$, and there is no BD ghost \cite{Hinterbichler:2012cn}, so the line theory in some sense is a more faithful extension of the single field dRGT model.

Positivity bounds have been used to constrain the parameter space of massive spin-2 theories \cite{Cheung:2016yqr, deRham:2018qqo, Bellazzini:2017fep, Bonifacio:2016wcb, Bonifacio:2018vzv, Alberte:2019xfh, Alberte:2019zhd}. The existence of these bounds only rely on the weak assumption that the UV completion of the EFT satisfy fundamental properties of the S-matrix such as Lorentz invariance, unitarity, locality, crossing symmetry and analyticity, which allow us to derive dispersion relations that link the EFT amplitudes to dispersive integrals of the UV theory. Utilizing the forward dispersion relation and the optical theorem, {\it i.e.}, the absorptive part of the forward elastic amplitude being positive, one can prove the forward elastic positivity bounds \cite{Adams:2006sv}. In fact, any $t$ ($s,t,u$ being the standard Mandelstam variables) derivatives of the absorptive part of the amplitude is also positive even slightly away from the forward limit, and this can be used to derive an infinite number of $t$ derivative positivity bounds \cite{deRham:2017avq, deRham:2017zjm} (also see \cite{AHH} and for non-forward bounds without $t$ derivatives see \cite{Manohar:2008tc, Bellazzini:2016xrt}). Making further judicial use of the partial wave extension and the full crossing symmetry, we can derive a set of new positivity bounds, which improve the $t$ derivative bounds of \cite{deRham:2017avq} and can bound the Wilson coefficients from both sides \cite{Tolley:2020gtv} (see also \cite{Bellazzini:2020cot, Caron-Huot:2020cmc, Guerrieri:2020bto}). Also, positivity bounds have been used to constrain the Standard Model Effective Field Theory \cite{Zhang:2018shp, Bi:2019phv, Yamashita:2020gtt, Remmen:2020vts, Remmen:2019cyz, Bellazzini:2017bkb, Bellazzini:2018paj, Remmen:2020uze, Zhang:2020jyn}. Absent any discovery of new particles at the LHC, the EFT approach has been gaining popularity in parametrizing new physics beyond the Standard Model. Indeed, positivity bounds have been shown to significantly reduce the size of the viable parameter space of this EFT \cite{Zhang:2018shp, Bi:2019phv, Yamashita:2020gtt, Remmen:2020vts, Remmen:2019cyz}. While providing important guidance for future experimental searches, they could also be used to test the fundamental properties of the S-matrix on collider experiments  \cite{Fuks:2020ujk, Gu:2020ldn}. In addition, a connection exists between the bounded parameter space and the possible new physical states  \cite{Zhang:2020jyn}.

The forward elastic positivity bounds can be applied to scattering amplitude $ab\to ab$, where arbitrary (or ``indefinite'') polarizations can be chosen for particle $a$ and particle $b$. In problems with multiple species of particles involved, we may also superimpose different particle species in addition to the superposition of different polarizations for the  external states, and get generalized elastic positivity bounds. That is,  we apply the positivity bounds to a linear combination of elastic {\it and inelastic} amplitudes
\beq
\label{ampMaibi0}
\mc{M}_{\ai\bi}=\sum_{a,b,c,d}\ai_{a}\bi_{b}\ai_{c}\bi_{d}\mc{M}_{ab\rightarrow cd}  ,
\eeq
with $\ai_a$ and $\bi_b$ being sets of arbitrary real constants, which may be viewed as the scattering amplitude for external state $|\ai\rangle=\sum_a \ai_a | a \rangle$ and $|\bi\rangle=\sum_a\bi_a |a\rangle$. If the particles are all of the same mass, it is straightforward to prove the forward positivity bounds for amplitude $\mc{M}_{\ai\bi}$. If the particles have different masses, however, care should be taken. First of all, for inelastic scatterings, the amplitude contains kinematical singularities even in the $t=0$ limit, which we need to regularize. In addition, notice that for an inelastic scattering between particles with different masses, $t=0$ is {\it not} the forward scattering limit, where crossing relations are mostly trivial even for particles with spin. Furthermore, in deriving the dispersion relation, we need crossing to map the left hand cut to the right hand cut to establish positivity. For an inelastic scattering amplitude with the kinematic singularities regularized, this introduces dependences on masses of the external particles in the dispersive integrand, because of which we can not use \eref{ampMaibi0} to establish strict positivity. However, for improved positivity bounds \cite{Bellazzini:2016xrt, deRham:2017imi} or for weakly coupled tree level positivity bounds, the integration starts around the cutoff of the EFT, which by the very validity of the EFT is much greater than the masses of low energy modes, so the mass dependences can be neglected and we still get generalized elastic positivity bounds, up to some corrections with an extra suppression of $\mc{O}(m^2/\Lambda_3^2)$.  All of these will be explained in detail in Section \ref{sec:gepb}.

The usual elastic positivity bounds have been applied to the bi-field cycle and line theory, and various Wilson coefficients of both theories have been constrained \cite{Alberte:2019xfh}. However, since the usual elastic positivity bounds do not access the information of inelastic processes, some coefficients are totally unconstrained. We apply generalized elastic positivity bounds to the bi-field cycle and line theory, and find that, for both theories, all the Wilson coefficients of the operators with the highest cutoff are now completely constrained to a finite region. Also, for the coefficients that are already constrained by the usual elastic bounds, the generalized bounds allow us to see how the constraints on these coefficients tighten for different choices of the coefficients that are unconstrained by the usual bounds.

The cycle and line theory may be considered as gravitational theories with diffeomorphism invariances that are broken by graviton potential terms. In the EFT context around the flat space, {\it a priori}, it may also be justified to consider interacting spin-2 theories with broken linearized diffeomorphism invariances. This is the case of pseudo-linear spin-2 theories \cite{Hinterbichler:2013eza}. Elastic positivity bounds have been applied to bi-field pseudo-linear spin-2 theories \cite{Alberte:2019zhd}, and it is found that for the Wilson coefficients that are constrained by the elastic positivity bounds, positivity requires their values to vanish. We now apply the generalized elastic positivity bounds to this theory, and find that we can exclude the whole parameter space of the theory, including the remaining Wilson coefficients unconstrained by the previous elastic bounds. Thus, we can completely rule out the bi-field pseudo-linear spin-2 theory to have a standard UV completion.
The simplicity of this theory also acts as a simple example to showcase the working mechanism of generalized elastic bounds, which will be discussed first in Section \ref{sec:pseudo}.

The remaining of the paper is organized as follows. In Section \ref{sec:gepb}, we set up some kinematic conventions for elastic and inelastic scatterings, review the kinematic singularities in inelastic scattering amplitudes and the method to regularize them, and derive generalized elastic positivity bounds. In Section \ref{sec:pseudo}, we introduce the bi-field pseudo-linear spin-2 theory, and apply generalized elastic positivity bounds to it, completely ruling out the theory to have an analytical UV completion. In Section \ref{sec:massspin2}, we introduce the bi-field cycle and line theory and then apply generalized elastic bounds to them; We find that all the Wilson coefficients in either theory are now constrained to finite regions. In Section \ref{sec:multifield}, we briefly consider generalizations of our arguments to cases where there are more than two dynamical spin-2 fields.

\section{Generalized elastic positivity bounds}

\label{sec:gepb}

Forward positivity bounds are usually applied to elastic scattering amplitudes $\mc{M}_{ab\rightarrow ab}$ from particle $a$ and particle $b$ to particle $a$ and particle $b$, where $a$ and $b$ are supplied with generic/indefinite polarizations. That is, the polarizations of $a$ and $b$ are arbitrary linear superpositions of the base polarizations. We can generalize this by also linearly superposing different particles to get the generalized elastic positivity bounds. For this, we consider a combination of (elastic and inelastic) amplitudes
\beq
\label{ampMaibi}
\mc{M}_{\ai\bi}=\sum_{a,b,c,d}\ai_{a}\bi_{b}\ai_{c}\bi_{d}\mc{M}_{ab\rightarrow cd}  ,
\eeq
with $\ai_a$ and $\bi_b$ being vectors of arbitrary real constants, which can be viewed as the scattering amplitude between state $|\ai\rangle=\sum_a \ai_a | a \rangle$ and $|\bi\rangle=\sum_a\bi_a |a\rangle$. In combining the different amplitudes $\mc{M}_{ab\rightarrow cd}$ in \eref{ampMaibi}, we let $\mc{M}_{ab\rightarrow cd}$ have the same values of the standard Mandelstam variables $s,t$. As we will see shortly, this necessarily means that the external momenta can be different for the different amplitudes, particularly the inelastic ones. So the   $|\ai\rangle$ and $|\bi\rangle$ states are not the usual one-particle asymptotic states of the scattering amplitude. Nevertheless, $\mc{M}_{\ai\bi}$ may be viewed as a transition amplitude in the broad sense, and it is an analytic function of $s,t$ because $\ai_{a},\bi_{b}$ are merely constants and $\mc{M}_{ab\rightarrow cd}$ are the standard scattering amplitudes with the usual analytic properties. For later convenience, we re-write this amplitude as follows
\beq
\label{ampMaibi2}
\mc{M}_{\ai\bi}=\sum_{i,j,k,l}\ai_{i}\bi_{j}\ai_{k}\bi_{l}\mc{M}_{ij\rightarrow kl}  ,
\eeq
where now the new summation indices $i,j,k,l$, which will still be referred to as particles, run over all the different species of particles and as well as their different polarizations. That is, particle $i$ has a definite polarization and other definite quantum numbers. If all the particles are of the same mass, it is straightforward to follow the usual steps to derive forward positivity bounds for these generalized elastic scattering amplitudes. When the masses are different, subtleties arise as we will see shortly. However,  we will show that for any EFT with a healthy hierarchy between the mass scale of the low energy modes and the EFT cutoff, the forward positivity bounds are still valid. Now, since amplitude $\mc{M}_{\ai\bi}$ contains information about inelastic scatterings, in addition to the elastic scatterings, it is conceivable that the positivity bounds on $\mc{M}_{\ai\bi}$ give rise to extra constraints on the Wilson coefficients compared to the usual elastic positivity bounds.

\subsection{Kinematics of inelastic scatterings}

One subtlety for inelastic scatterings is that the forward limit generically does not coincide with $t=0$. To see this, let us consider an inelastic scattering from particle $i$ and particle $j$ to particle $k$ and particle $l$.  In the center of mass frame, the momenta of the particles in the scattering can be parameterized as
\bal
&p_i^\mu=\(w_i,k_i\mathrm{sin}\thi_i,0,k_i\mathrm{cos}\thi_i\),~
p_j^\mu=\(w_j,k_i\mathrm{sin}\thi_j,0,k_i\mathrm{cos}\thi_j\) ,\\
&p_k^\mu=\(w_k,k_k\mathrm{sin}\thi_k,0,k_k\mathrm{cos}\thi_k\),~
p_l^\mu=\(w_l,k_k\mathrm{sin}\thi_l,0,k_k\mathrm{cos}\thi_l\)  ,
\eal
with
\bal
\label{wandk1}
w_i=\frac{s+m_i^2-m_j^2}{2\sqrt{s}},&~~w_j=\frac{s+m_j^2-m_i^2}{2\sqrt{s}} ,\\
w_k=\frac{s+m_k^2-m_l^2}{2\sqrt{s}},&~~w_l=\frac{s+m_l^2-m_k^2}{2\sqrt{s}}  ,\\
\label{wandk2}
k_i=\sqrt{\f{\mc{S}_{ij}}{4s}},~~~~~~~~~&~~k_k=\sqrt{\f{\mc{S}_{kl}}{4s}}  ,
\eal
where $m_i,m_j,m_k,m_l$ are the masses of the interacting particles and we have defined the usual Mandelstam variables
\beq
s=-(p_i+p_j)^2,~t=-(p_i-p_k)^2,~u=-(p_i-p_l)^2  ,
\eeq
and also defined
\be
\label{Sijdef}
\mc{S}_{ij}=\[s-(m_i+m_j)^2\] \[s-(m_i-m_j)^2\]   .
\ee
Without loss of generality, we can choose the scattering angles such that $\thi_i=0$, $\thi_j=\pi$, $\thi_k=\thi$, $\thi_l=\thi+\pi$, then we have
\beq
\label{cosexpr}
\mathrm{cos}~ \theta=\frac{2 w_i w_k-m_i^2-m_k^2+t}{2k_i k_k}  .
\eeq
We see that in inelastic scatterings generically the forward scattering $\thi=0$ {\it does not} correspond to $t=0$. (When the $i$ and $k$ particle, as well as the $j$ and $l$ particle, have identical mass $m_i=m_k, m_j=m_l$, we have $\cos \thi-1 ={2st}/{[s-{(m_i+m_j)^2}][s+(m_i-m_j)^2]}$, in which case the forward limit becomes the same as the $t=0$ limit.) This will pose as a difficulty to generalize positivity bounds to include generic inelastic scatterings, as crossing relations are more complicated in non-forward scatterings. However,  as we shall see, as long as there is a healthy hierarchy between the cutoff and the mass scale of the EFT, possible discrepancies are higher order effects on the right hand side of the dispersion relation and can be neglected.

A convenient set of polarizations to use in forward positivity bounds are in the linear basis.  For a massive spin-1 particle with momentum $p^{\mu}=\left(w, k \sin \theta, 0, k \cos \theta\right)$, we have
\beq
\epi_\mu^1=(0,\cos \thi,0, -\sin \thi),~
\epi_\mu^2=(0,0,1,0),~
\epi_\mu^3=(k,w\sin\thi,0,w\cos\thi)/m ,
\eeq
which satisfy $p^{\mu} \epsilon_{\mu}^{i}=0$ and normalizaiton $\eta^{\mu\nu}\epsilon_{\mu}^{i} \epsilon^{j}_{\nu}=\delta^{i j}$. The massive spin-2 polarization tensors can be constructed with the spin-1 polarization vectors, that is, we have
\bal
&\epi_{\mu\nu}^{1}=\frac{1}{\sqrt{2}}\(\epi_\mu^1\epi_\nu^2-\epi_\mu^2\epi_\nu^1\),
~\epi_{\mu\nu}^{2}=\frac{1}{\sqrt{2}}\(\epi_\mu^1\epi_\nu^2+\epi_\mu^2\epi_\nu^1\),\\
&\epi_{\mu\nu}^{3}=\frac{i}{\sqrt{2}}\(\epi_\mu^1\epi_\nu^3+\epi_\mu^3\epi_\nu^1\),
~\epi_{\mu\nu}^{4}=\frac{i}{\sqrt{2}}\(\epi_\mu^2\epi_\nu^3+\epi_\mu^3\epi_\nu^2\),\\
&\epi_{\mu\nu}^{5}=\sqrt{\frac{3}{2}}\(\epi_\mu^3\epi_\nu^3-\frac{1}{3}\(\ei_{\mu\nu}+\frac{p_\mu p_\nu}{m^2}\)\).
\eal
which satisfy $k^{\mu} \epsilon_{\mu \nu}^{i}=\eta^{\mu\nu}\epsilon^{i}_{\mu\nu}=0$ and normalization $\eta^{\mu\rho}\eta^{\nu\si}\epsilon_{\mu \nu}^{i} \epsilon^{j}_{\rho \si}=\delta^{i j}$.

\subsection{Singularities in inelastic scatterings}

The analyticity of the scattering amplitude, upon using the Cauchy's integral formula and the Froissart-Martin bound for UV amplitudes, gives rise to dispersion relations, which link the IR and the UV physics and provide an important means to probe non-perturbative information of quantum field theory. Before deriving the dispersion relation, let us review the singularity structure of a general inelastic scattering amplitude $\mc{M}_{ijkl}$.

{\bf Physical singularities}: Tree level exchanges give rise to simple poles at $s=m_n^2$ with $n$ denoting the masses of all possible exchange channels, and loop level amplitudes give rise to branch cuts starting from the lowest threshold $s=s_0$ to infinity. From the $s \leftrightarrow u$ crossing, generally, there are also poles at $u=m_n^2$ and branch cuts from $u=s_0$ to infinity. In complex $s$ plane, the locations of the $u$ channel singularities depend on the masses of the specific scattering process, $s=m_i^2+m_j^2+m_k^2+m_l^2-t-u$. Those are called physical singularities as they are linked to the physical process or the physical spectrum of the theory and they already appear in amplitudes for scalars.

{\bf Kinematical singularities}: For scatterings of particles with spin, extra so-called kinematical singularities arise in the amplitude \cite{Cohen-Tannoudji:1968lnm}. To see where these singularities originate from, we note that the $w_i$ and $k_i$ from \eref{wandk1} to \eref{wandk2} and also $\cos\thi$ and $\sin\thi$ become singular at various places and the polarizations, which are part of the amplitude, are built out of these singular quantities. Note that for inelastic scatterings $\cos \theta$ has poles even when $t=0$, as can be seen from \eref{cosexpr}. In the following we list the relevant kinematical singularities for an inelastic scattering amplitude $\mc{M}_{ijkl}$ from particle $i$ and particle $j$ to particle $k$ and particle $l$, and specify their regularization methods.
\begin{itemize}

\item Branch points at $s=(m_i+m_j)^2,~(m_i-m_j)^2,~(m_k+m_l)^2,~(m_k-m_l)^2$, where $m_i, m_j,m_k,m_l$ are the masses of the external particles. They come from the square roots in $k_i$ and $k_k$ of \eref{wandk2}. 

To remove these branch points, we can simply superimpose the amplitude with different signs of $k_i$ and $k_k$ to get rid of the terms with odd powers of $k_i$ and $k_k$ in the amplitude. Specifically, we can define
\be
\label{barM}
\hat{\mc{M}}_{ijkl}=
\begin{cases}
\frac{1}{4}
[\mc{M}_{ijkl}(k_i,k_k)+\mc{M}_{ijkl}(-k_i,k_k)+\mc{M}_{ijkl}(k_i,-k_k)+\mc{M}_{ijkl}(-k_i,-k_k)],&{\rm if} ~k_i\neq k_k
\\
\frac{1}{2}
[\{\mc{M}_{ijkl}(k_i,k_k)+\mc{M}_{ijkl}(-k_i,k_k)],&{\rm if}~ k_i=k_k
\end{cases}
\ee
and use $\hat{\mc{M}}_{ijkl}$ instead.

\item  Poles at $s=(m_i+m_j)^2,~(m_i-m_j)^2,~(m_k+m_l)^2,~(m_k-m_l)^2$. These originate from the singularities of $\cos\thi$ and $\sin\thi$ (and also $\cos\f{\thi}2$ and $\sin\f{\thi}2$ when fermions are involved). As mentioned, $\cos\thi$ contains poles even when $t = 0$ for inelastic scatterings. The order of these kinematical poles are related to the spins of the particles. For example, in the case of massive spin-2 particle scatterings, each pole is at most second-order. Generally the order of pole $s=(m_i+m_j)^2$ and pole $s=(m_i-m_j)^2$ is at most $(S_i+S_j)/2$, and the order of pole $s=(m_k+m_l)^2$ and pole $s=(m_k-m_l)^2$ is at most $(S_k+S_l)/2$. Intuitively, one may count the order of the poles by counting the powers of $\cos\thi$ and $\sin\thi$. Since $\cos\thi$ and $\sin\thi$ come from rotating some standard polarizations, the powers of $\cos\thi$ and $\sin\thi$ reflect the spins of the interacting particles, which thus determine the maximum orders of the poles.

To remove all the kinematical poles, we can multiply the amplitude by an overall factor to define another modified amplitude
\beq
\label{SSpower}
\bar{\mc{M}}_{ijkl}= \mc{S}_{ij}^{ (S_i+S_j)/2}\mc{S}_{kl}^{ (S_k+S_l)/2} \hat{\mc{M}}_{ijkl}  ,
\eeq
where $S_i$ and $S_j$ are the spins of particle $i$ and particle $j$ respectively and $\mc{S}_{ij}$ is defined in \eref{Sijdef}.

\end{itemize}

\subsection{Generalized elastic positivity bounds}

With all the kinematical singularities removed, we can use  $\bar{\mc{M}}_{ijkl}$ to derive a dispersion relation. Here we will consider the $t=0$ limit of the amplitude $\bar{\mc{M}}_{ijkl}(s)=\bar{\mc{M}}_{ijkl}(s,t=0)$. Note that this is not necessarily the forward limit $\thi=0$ for a generic inelastic scattering. However, as we shall see, on the right hand side of the dispersion relation with the low energy part of the integral subtracted or for the tree level dispersion relation in the case of a weakly coupled theory, we can approximate the $t=0$ limit with the forward limit $\thi=0$, and thus the standard forward positivity argument can apply.

To see this, following the usual steps (see, {\it e.g.}, \cite{deRham:2017zjm}), by the analyticity of the complex $s$ plane and the Froissart-Martin bound, we can get
\beq
\bar{\mc{M}}_{ijkl}(s)=(\mathrm{Physical~poles})+\int_{s_0}^{\infty} \frac{\d \mu}{2i\pi} \(\frac{\mathrm{Disc}\bar{\mc{M}}_{ijkl}(\mu)}{\mu-s}+\frac{\mathrm{Disc}\bar{\mc{M}}_{ilkj}(\mu)}{\mu-\Delta_{ijkl}+s}\)  ,
\eeq
where ``$(\mathrm{Physical~poles})$'' denotes the terms involving the physical poles, the discontinuity ``Disc'' is defined as $\mathrm{Disc} \bar{\mc{M}}_{ijkl}(\mu) =\bar{\mc{M}}_{ijkl}(\mu+i\varepsilon)-\bar{\mc{M}}_{ijkl}(\mu-i\varepsilon)$ and $\Delta_{ijkl}$ is the sum of the mass squared of the 4 external particles $\Delta_{ijkl}=m_i^2+m_j^2+m_k^2+m_l^2$. For a heathy EFT where its cutoff $\Lambda$ is much greater than the mass scale of the low energy modes $\mc{O}(\Delta_{ijkl}^{1/2})$, we can compute the amplitude to a desired accuracy up to energy scale $\sqrt{s}<\epi\Lambda$. Thus, we can subtract out the low energy part of the dispersive integral up to $\epi\Lambda$ with $\epi\lesssim 1$ \cite{Bellazzini:2016xrt, deRham:2017imi} and get
\beq
\bar{\mc{M}}^{\epi\Lambda}_{ijkl}(s)=(\mathrm{Physical~poles})+\int_{(\epi\Li)^2}^{\infty} \frac{\d \mu}{2i\pi} \(\frac{\mathrm{Disc}\bar{\mc{M}}_{ijkl}(\mu)}{\mu-s}+\frac{\mathrm{Disc}\bar{\mc{M}}_{ilkj}(\mu)}{\mu-\Delta_{ijkl}+s}\)  .
\eeq
For an EFT that is weakly coupled, one can derive the dispersion relation with the tree level amplitude, for which case, $\epi\Lambda$ will be the mass scale of the first particle that is not captured in the EFT, around the scale of cutoff $\Lambda$, and this is the scenario we are assuming for the spin-2 EFTs studied in the following sections.

Let $n$ be the order $s$ in $\mc{S}_{ij}^{(S_i+S_j)/2}\mc{S}_{kl}^{(S_k+S_l)/2}$ in \eref{SSpower}. We can perform $n+2$-order $s$ derivative and evaluate it at $s=\Delta_{ijkl}/2$:
\bal
f_{ijkl}&=\frac{1}{(n+2)!}\frac{\d^{n+2}}{\d s^{n+2}}\[\bar{\mc{M}}^{\epi\Lambda}_{ijkl}(s,0)-(\mathrm{Physical~poles})\]_{s\rightarrow \Delta_{ijkl}/2}\\
\label{dispersion}
&=\int_{(\epi\Li)^2}^{\infty} \frac{\d \mu}{2i\pi} \(\frac{\mathrm{Disc}\bar{\mc{M}}_{ijkl}(\mu)}{(\mu-\Delta_{ijkl}/2)^{n+3}}+\frac{\mathrm{Disc}\bar{\mc{M}}_{ilkj}(\mu)}{(\mu-\Delta_{ijkl}/2)^{n+3}}\) .
\eal
Now, since $\mu \geq (\epi\Li)^2 \gg \Delta_{ijkl}$, we can neglect $\Delta_{ijkl}$ in the denominator of the dispersive integral. Also, for the amplitude in the dispersive integral $\bar{\mc{M}}_{ijkl}(\mu)$, since $\mu \gg \Delta_{ijkl}$, the $t=0$ limit becomes the limit of $\cos \theta|_{s\to \mu}=1$, {\it i.e.,} the forward scatting limit of $\bar{\mc{M}}_{ijkl}(\mu)$, up to higher order corrections, as can be seen from \eref{cosexpr}. In the forward limit, the $s-u$ crossing for ${\mc{M}}_{ijkl}(\mu)$ is trivial, while the $s-u$ crossing in the limit $t=0$ would be rather complicated. Also, because of $\mu \gg \Delta_{ijkl}$, we have $\bar{\mc{M}}_{ijkl} \simeq \mu^{14} \mc{M}_{ijkl}$. Thus the $s-u$ crossing for $\bar{\mc{M}}_{ijkl}$ also becomes trivial and we finally get
\beq
f_{\ai\bi}=\sum_{i,j,k,l}\ai_i\bi_j\ai_k\bi_l f_{ijkl}=\int_{(\epi\Li)^2}^{\infty} \frac{\d \mu}{\pi} \frac{\f1i\mathrm{Disc} \bar{\mc{M}}_{\ai\bi}(\mu)}{\mu^{n+3}}  ,
\eeq
where we emphasize that $i,j,k,l$ run over all the different particles and as well as the different polarizations, $\ai_i$ and $\bi_j$ are sets of arbitrary constants and we have defined
\be
\bar{\mc{M}}_{\ai\bi}(\mu) = \sum_{i,j,k,l}\ai_i\bi_j\ai_k\bi_l\bar{\mc{M}}_{ijkl}(\mu)  .
\ee
Note that since we have taken the limit $\mu \geq (\epi \Li)^2\gg \Delta_{ijkl}$ on the right side of \eref{dispersion}, strictly speaking, we should also only keep the leading contributions for the whole dispersive integral after the integration is carried out, and our dispersive integrals here should be understood in this sense. This means that we should also take the limit $E\ll \Lambda$ (with $ E^2\sim \Delta_{ijkl},s$) on the left side of \eref{dispersion}.  In other words, we should only keep the leading order EFT amplitude in $E/\Lambda$ on the left hand side.

Now, Hermitian analyticity and the generalized optical theorem implies that
\bal
 \f1i\mathrm{Disc} \bar{\mc{M}}_{\ai\bi}
&=\f1i\sum_{i,j,k,l}\ai_i\bi_j\ai_k\bi_l \(\bar{\mc{M}}_{ijkl}-\bar{\mc{M}}^*_{klij}\)
= \sum_{i,j,k,l}\sum_{[X]}\(\ai_i\bi_j \bar{\mc{M}}_{ij\rightarrow X}\)\(\ai_k\bi_l \bar{\mc{M}}_{kl\rightarrow X}\)^*>0  ,
\eal
where $[X]$ denotes summation over intermediate states along with their phase space integration. Therefore, we arrive at the generalized elastic positivity bounds
\be
f_{\ai\bi} = \sum_{i,j,k,l}\ai_i\bi_j\ai_k\bi_l f_{ijkl}> 0, ~~{\rm for~any~real}~\ai_i,\bi_i  .
\ee
In the above, we take $\ai_i$ and $\bi_j$ to be real constants. If we let $\ai_i$ and $\bi_j$ be complex and replace $\sum_{i,j,k,l}\ai_i\bi_j\ai_k\bi_l$ with $\sum_{i,j,k,l}\ai_i\bi_j\ai^*_k\bi^*_l$, the positivity argument still goes through. However, empirically, we find that extending to the complex domain does not enhance the positivity bounds, at least not significantly. (In fact, this is rigorously true in the massless limit, because in that limit we do not have kinematic singularities and by choosing appropriate polarizations for the external states we have $f_{ijkl}=f_{ilkj}=f_{klij}$, which enforces the same symmetries for the indices on $\ai$ and $\bi$. Then if we let $\ai_i=u_i+i v_i$ and $\bi_i=r_i+i s_i$,  $\ai_i\bi_j\ai^*_k\bi^*_l$ reduces to
\be
 \ai_i\bi_j\ai^*_k\bi^*_l = u_i r_j u_k r_l +   u_i s_j u_k s_l  + v_i r_j v_k r_l +u_i s_j u_k s_l  .
\ee
which is a positive sum of real tensors of the $\ai\bi\ai\bi$ form, meaning that it is sufficient to only consider real parameters $\ai_i\bi_j\ai_k\bi_l$ when mixing different modes.)

\section{Positivity on interacting pseudo-linear spin-2 theory}
\label{sec:pseudo}

Let us start with a simpler theory, interacting pseudo-linear spin-2 theory. (Here interacting refers to interaction between multiple field species, rather than nonlinear interactions that can be there in the case of a single field species.) It is easy to see from this simple case why mixing different particle species can give rise to new bounds on the Wilson coefficients unconstrained by considering the usual elastic positivity bounds. For simplicity, we will focus on the bi-field case in this section. Generalizations to multiple fields will be briefly considered in Section \ref{sec:multifield}.

\subsection{Interacting pseudo-linear theory}

As mentioned in the introduction, (single field) pseudo-linear spin-2 theory is a simple generalization of the linear Fierz-Pauli action with nonlinear interactions that do not introduce extra Ostrogradski degrees of freedom but without the nonlinear interactions of the Einstein-Hilbert term \cite{Hinterbichler:2013eza}. These interactions are the leading terms in the Wilsonian effective action, giving rise to the highest cutoff for the EFT. Indeed, the decoupling limit of pseudo-linear spin-2 theory is the same as that of the dRGT massive gravity. Thus, pseudo-linear spin-2 theory is a simplified $\Lambda_3$ spin-2 theory and can be taken as a toy model of $\Lambda_3$ massive gravity, if by itself is not phenomenologically viable.

The pseudo-linear interactions include nonlinear terms invariant under linearized diffeomorphisms
\be
h_{\mu\nu}\to h_{\mu\nu}+\pd_\mu\xi_\nu + \pd_\nu\xi_\mu   ,
\ee
where in this section, different from the other sections, $h_{\mu\nu}$ is the perturbative metric $h_{\mu\nu}=M(g_{\mu\nu}-\eta_{\mu\nu})$, with $\eta_{\mu\nu}$ being the Minkowski metric and $M$ being a normalization mass scale to be specified later. These are the leading non-trivial terms from the Lovelock action. (When expanding each Lovelock term around the flat background, the leading term is a total derivative and thus is trivial; the next leading term is non-trivial, giving rise to the pseudo-linear term invariant under linearized diffeomorphisms.) The pseudo-linear interactions also include leading dRGT potential terms and special pseudo-linear derivative terms that break linearized diffeomorphisms. All the three kinds of terms are structurally similar, and a generic term with $d$ derivatives and $n$-th order in $h^\mu{}_\nu$ in $D$ dimension can be written as
\bal
\label{Ldn}
\mc{L}_{d,n}&\propto  \pd^{[\mu_1}\pd_{\mu_1}  h^{\mu_2}{}_{\mu_2} ...\pd^{\mu_{d-1}}\pd_{\mu_{d-1}}  h^{\mu_d}{}_{\mu_d}  h^{\mu_{d+1}}{}_{\mu_{d+1}}...h^{\mu_{\small n-\f{d}2}}{}_{ \mu_{\small n-\f{d}2}}\dd^{\mu_{\small n-\f{d}2+1}}{}_{\mu_{\small n-\f{d}2+1}}...\dd^{\mu_D]}{}_{\mu_D}
\\
 &\propto  \epi_{\mu_1...\mu_D}\epi^{\nu_1...\nu_D} \pd^{\mu_1}\pd_{\nu_1}  h^{\mu_2}{}_{\nu_2} ...\pd^{\mu_{d-1}}\pd_{\nu_{d-1}}  h^{\mu_d}{}_{\nu_d}  h^{\mu_{d+1}}{}_{\nu_{d+1}}...h^{\mu_{\small n-\f{d}2}}{}_{ \nu_{\small n-\f{d}2}}\dd^{\mu_{\small n-\f{d}2+1}}{}_{\nu_{\small n-\f{d}2+1}}...\dd^{\mu_D}{}_{\nu_D}   ,
 \eal
where $h^\mu{}_{\nu}=\eta^{\mu\ri}h_{\ri\nu}$ and $\epi_{\mu_1...\mu_D}$ is the Minkowski space Levi-Civita tensor in $D$ dimensions. In this notation, $\mc{L}_{2,2}$ is the linearized Einstein-Hilbert Lagrangian, $\mc{L}_{d,\f{d}{2}+1}$ is the pseudo-linear term from a general Lovelock term, $\mc{L}_{0,2}$ is the Fierz-Pauli mass term, and $\mc{L}_{0,n}$ is the pseudo-linear term from the corresponding dRGT potential term. We will focus on 4D in this paper, in which case  there are only one special pseudo-linear derivative term (with two derivatives), two leading dRGT potential terms and no contribution from the Lovelock terms. The positivity bounds on (single field) pseudo-linear spin-2 theory have been considered in \cite{Bonifacio:2016wcb}, which excludes the whole parameter space of the single field theory to have a standard UV completion that satisfies the fundamental principles of the S-matrix.

The generalization of the pseudo-linear theory to multiple fields is straightforward, simply replacing some of the $h^\mu{}_{\nu}$ in \eref{Ldn} with extra spin-2 fields. In this section, for simplicity, we shall only consider two such fields, denoted as $h_{\mu\nu}$ and $f_{\mu\nu}$ respectively. Adopting the notation of \cite{Alberte:2019zhd}, the general bi-field pseudo-linear Lagrangian is given by \cite{Bonifacio:2018van},

\bal
g_*^2\mc{L}_{\rm pseudo} = &~\mc{L}_{\mathrm{FP}}(h,m_1)+\frac{a_1}{2M_1}\epi\epi(\pd^2 h)hh
+\frac{m_1^2}{4}
\[\frac{\ki_3^{(1)}}{M_1}\epi \epi Ihhh+\frac{\ki_4^{(1)}}{M_1^2}\epi \epi hhhh\]\nn
&+\mc{L}_{\mathrm{FP}}(f,m_2)+\frac{a_2}{2M_2}\epi\epi(\pd^2 f)ff +\frac{m_2^2}{4}
\[\frac{\ki_3^{(2)}}{M_2}\epi \epi Ifff+\frac{\ki_4^{(2)}}{M_2^2}\epi \epi ffff\]\nn
&+\frac{a_3}{2M_1}\epi\epi(\pd^2 h)hf+\frac{a_4}{2M_2}\epi\epi(\pd^2 h)ff+\frac{a_5}{2M_2}\epi\epi(\pd^2 f)fh+\frac{a_6}{2M_1}\epi\epi(\pd^2 f)hh\nn
&+\frac{m_2^2 c_1}{2M_1}\epi \epi Ihhf+\frac{m_2^2 c_2}{2M_2}\epi \epi Iffh+\frac{m_2^2\li}{2M_1 M_2}\epi \epi hhff+\frac{m_2^2 d_1}{4M_1^2}\epi \epi hhhf+\frac{m_2^2 d_2}{4M_2^2}\epi \epi hfff +...   ,
\label{pseudo3}
\eal
where the Fierz--Pauli Lagrangian is canonically normalized
\be
\label{FPlinear}
g_*^2\mc{L}_{\mathrm{FP}}(h,m_1) = -\frac{1}{2} \partial_{\lambda} h_{\mu \nu} \partial^{\lambda} h^{\mu \nu}+\partial_{\mu} h_{\nu \lambda} \partial^{\nu} h^{\mu \lambda}-\partial_{\mu} h^{\mu \nu} \partial_{\nu} h^\ri{}_\ri+\frac{1}{2} \partial_{\lambda} h^\ri{}_\ri \partial^{\lambda} h^\ri{}_\ri-\frac{1}{2} m_1^{2}\left(h_{\mu \nu} h^{\mu \nu}-(h^\ri{}_\ri)^{2}\right) ,
\ee
and $...$ stands for subleading higher derivative terms.
For the interactions terms, we have used the short hand notation for the double Levi-Civita contractions
\be
\epi \epi ABCD  \equiv -\epi_{\mu\nu\rho\si}\epi^{\ai\bi\gi\di}A^\mu{}_\ai B^\nu{}_\bi C^\rho{}_\gi D^\si{}_\di   .
\ee
For example, we have $\epi\epi(\pd^2 h)hh = -\epi_{\mu\nu\rho\si}\epi^{\ai\bi\gi\di}\pd^\mu\pd_\ai h^\nu{}_\bi h^\rho{}_\gi h^\si{}_\di$, $\epi \epi Ihhh=-\epi_{\mu\nu\rho\si}\epi^{\ai\bi\gi\di}\dd^\mu{}_\ai h^\nu{}_\bi h^\rho{}_\gi h^\si{}_\di$, etc. $m_1$ and $m_2$ are the masses of the two massive gravitons $h_{\mu\nu}$ and $f_{\mu\nu}$ respectively, which are assumed to not have a big hierarchy between them $m_1\sim m_2\equiv m$. $M_1$ and $M_2$ are the nonlinearity scale of $h_{\mu\nu}$ and $f_{\mu\nu}$ and are also assumed to not have a big hierarchy between them $M_1\sim M_2\equiv M$. For a valid EFT, we assume $M_1,M_2\gg m_1,m_2$. While $M$ is the strong coupling scale of the helicity-2 modes,  the real cutoff of the theory is at $\Lambda_3=(m^2M)^{1/3}$ \cite{Alberte:2019zhd}, due to the lower strong coupling scale of the helicity-0 modes. $a_i,c_i,d_i,\ki^{(1)}_3,\ki^{(1)}_4,\ki^{(2)}_3,\ki^{(2)}_4,\li$ are dimensionless Wilson coefficients that are to be constrained by the positivity bounds. We have also introduced weak coupling $g_*^2\ll 1$, which suppresses the loop amplitudes and allows us to use the tree level positivity bounds. Indeed, for a standard weakly coupled UV standard to exist, the improved positivity bounds implies that $g_*^2\ll m^2/\Lambda^2\ll 1$ \cite{Bellazzini:2017fep, deRham:2017xox}.

\subsection{Exclusion by positivity}

In \cite{Alberte:2019zhd}, by applying the positivity bounds for forward elastic process $hh\rightarrow hh$ and $ff\rightarrow ff$, inequalities on the Wilson coefficients going both directions can be found and it is concluded that $\ki_{3,4}^{(1,2)}=0$, $a_i=0$, $c_i=0$. Also, by applying the positivity bounds for forward elastic process $hf\rightarrow hf$, one also finds that $\li=0$. So, to satisfy the forward elastic positivity bounds without mixing $h_{\mu\nu}$ and $f_{\mu\nu}$, we are left with the leading Lagrangian
\beq
\label{leftLpseudo}
g_*^2\mc{L}_{\rm pseudo} = \mc{L}_{\mathrm{FP}}(h,m_1)+ \mc{L}_{\mathrm{FP}}(f,m_2)+\frac{m_1^2 d_1}{4M_1^2}\epi \epi hhhf+\frac{m_2^2 d_2}{4M_2^2}\epi \epi hfff +...   .
\eeq
Obviously, the theory is formally the same under exchanging
\beq
\label{pseudosym}
h\leftrightarrow f,~m_1\leftrightarrow m_2,~M_1\leftrightarrow M_2,~ d_1 \leftrightarrow d_2   .
\eeq
Making use of this exchange symmetry we only need to calculate half of the all $16$ scattering amplitudes. In the following, we will show that generalized elastic positivity bounds also require the remaining $d_1$ and $d_2$ coefficients to vanish. Therefore, bi-field pseudo linear spin-2 theories do not have a standard Wilsonian UV completion, just like the single field pseudo-linear theory \cite{Bonifacio:2016wcb}.

As discussed in the previous section, generalized elastic positivity bounds can be viewed as bounds on scattering amplitudes between superpositions of different particle species, utilizing additionally the information from inelastic scattering amplitudes. Different from forward elastic scatterings for the same particle, inelastic scattering amplitudes have kinematic singularities even in the forward limit. We shall regularize the kinematic branch points as prescribed in \eref{barM}. For the kinematic poles, we could also regularize them as prescribed in \eref{SSpower} for each inelastic amplitudes and keep the original amplitudes for the elastic amplitudes. However, we find it convenient to simply multiply a single sufficient overall factor to regularize the poles for all the different amplitudes, compensated with an appropriate number of extra $s$ derivatives in the positivity bounds. To identity this overall factor, we note that the elastic $hh\to hh$ and $hf\to hf$ amplitude do not have any kinematic pole; for the $hh\to hf$, $hh\to fh$, $hf\to hh$ and $fh\to hh$ amplitude we need to multiply a factor of $s^2(s-4m_1^2)^2(s-(m_1+m_2)^2)^2(s-(m_1-m_2)^2)^2$; for the $hf\to fh$ amplitude we need multiply a factor of $(s-(m_1+m_2)^2)^4(s-(m_1-m_2)^2)^4$; for the $hh\to ff$ amplitude we need to multiply a factor of $s^2(s-4m_1^2)^2(s-4m_2^2)^2$. The kinematic poles of the rest amplitudes can be obtained by the formal symmetry of theory in exchanging $h$ and $f$. Therefore, the overall factor we need is
\beq
\Gi(s)=s^2\(s-4m_1^2\)^2\(s-4m_2^2\)^2\[s-(m_1+m_2)^2\]^4\[s-(m_1-m_2)^2\]^4  .
\eeq
For the pseudo-linear theory in this section, a simpler regularization factor would also be sufficient, but this factor is universal in the sense that it also works for other interacting spin-2 theories in the following sections. Then, we can compute the amplitude between state $|\ai\rangle$ and $|\bi\rangle$
\be
|\ai\rangle=\sum_{i=1}^5\ai_i |h,\epi^i\rangle +\sum_{i=6}^{10} \ai_i |f,\epi^{i-5}\rangle
,~~~
|\bi\rangle=\sum_{i=1}^5\bi_i |h,\epi^i\rangle +\sum_{i=6}^{10} \bi_i |f,\epi^{i-5}\rangle   ,
\ee
where $\ai_i$ and $\bi_i$ are arbitrary real constants and we have mixed different polarizations $\epi^i_{\mu\nu}$ as well as different particle species $h_{\mu\nu}$ and $f_{\mu\nu}$. Following the prescription above, we can obtain the following positivity bound for the regularized tree level amplitude $\bar{\mc{M}}_{\ai\bi}(s,0)$
\bal
f_{\ai\bi} =~&  \frac{1}{16!}\frac{\d^{16}}{\d s^{16}}\sum_{i,j,k,l}\ai_i\bi_j\ai_k\bi_l\[\bar{\mc{M}}_{ijkl}(s,0)-(\mathrm{Physical~poles})\]_{s\rightarrow \Delta_{ijkl}/2} > 0\\
=~&-\frac{1}{24m_1^2 m_2^2 M_1^2}d_1\cdot \nn
&\Big(6 \sqrt{3}  \ai _1 \ai _{10} \bi _3^2 m_1^2+6  \ai _5 \ai _{10} \bi _3^2 m_1^2+8  \ai _5 \ai _{10} \bi _5^2 m_1^2-6 \sqrt{3}  \ai _2 \ai _{10} \bi _3 \bi _4 m_1^2-24  \ai _1 \ai _{10} \bi _1 \bi _5 m_1^2\nn
+&12  \ai _2 \ai _{10} \bi _2 \bi _5 m_1^2+6 \sqrt{3}  \ai _3^2 \bi _1 \bi _{10} m_1^2-24  \ai _1 \ai _5 \bi _1 \bi _{10} m_1^2-6 \sqrt{3}  \ai _3 \ai _4 \bi _2 \bi _{10} m_1^2+12  \ai _2 \ai _5 \bi _2 \bi _{10} m_1^2\nn
+&6  \ai _3^2 \bi _5 \bi _{10} m_1^2+8  \ai _5^2 \bi _5 \bi _{10} m_1^2+12  \ai _3 \ai _8 \bi _5^2 m_1 m_2+9  \ai _4 \ai _8 \bi _3 \bi _4 m_1 m_2+9  \ai _3 \ai _9 \bi _3 \bi _4 m_1 m_2\nn
+&12 \sqrt{3}  \ai _3 \ai _8 \bi _1 \bi _5 m_1m_2-6 \sqrt{3}  \ai _4 \ai _8 \bi _2 \bi _5 m_1 m_2-6 \sqrt{3}  \ai _3 \ai _9 \bi _2 \bi _5 m_1 m_2+12  \ai _5^2 \bi _3 \bi _8 m_1 m_2\nn
+&12 \sqrt{3}  \ai _1 \ai _5 \bi _3 \bi _8 m_1 m_2+9  \ai _3 \ai _4 \bi _4 \bi _8 m_1 m_2-6 \sqrt{3}  \ai _2 \ai _5 \bi _4 \bi _8 m_1 m_2+9  \ai _3 \ai _4 \bi _3 \bi _9 m_1 m_2\nn
-&6 \sqrt{3}  \ai _2 \ai _5 \bi _3 \bi _9 m_1 m_2+6 \sqrt{3}  \ai _5 \ai _6 \bi _3^2 m_2^2 +6  \ai _5 \ai _{10} \bi _3^2 m_2^2+8  \ai _5 \ai _{10} \bi _5^2 m_2^2 -6 \sqrt{3}  \ai _5 \ai _7 \bi _3 \bi _4m_2^2\nn
-&24  \ai _5 \ai _6 \bi _1 \bi _5m_2^2 +12  \ai _5 \ai _7 \bi _2 \bi _5m_2^2 +6 \sqrt{3}  \ai _3^2 \bi _5 \bi _6 m_2^2-24  \ai _1 \ai _5 \bi _5 \bi _6m_2^2-6 \sqrt{3}  \ai _3 \ai _4 \bi _5 \bi _7m_2^2\nn
+&12  \ai _2 \ai _5 \bi _5 \bi _7m_2^2 +6  \ai _3^2 \bi _5 \bi _{10}m_2^2 +8  \ai _5^2 \bi _5 \bi _{10} m_2^2
\Big)\nn
&-\frac{1}{24m_1^2 m_2^2 M_2^2}d_2\cdot\nn
&\Big(6 \sqrt{3}  \ai _1 \ai _{10} \bi _8^2 m_1^2+6  \ai _5 \ai _{10} \bi _8^2 m_1^2+8  \ai _5 \ai _{10} \bi _{10}^2 m_1^2-6 \sqrt{3}  \ai _2 \ai _{10} \bi _8 \bi _9 m_1^2+6 \sqrt{3}  \ai _8^2 \bi _1 \bi _{10} m_1^2\nn
-&24  \ai _6 \ai _{10} \bi _1 \bi _{10} m_1^2-6 \sqrt{3}  \ai _8 \ai _9 \bi _2 \bi _{10} m_1^2+12  \ai _7 \ai _{10} \bi _2 \bi _{10} m_1^2+6  \ai _8^2 \bi _5 \bi _{10} m_1^2+8  \ai _{10}^2 \bi _5 \bi _{10} m_1^2\nn
-&24  \ai _1 \ai _{10} \bi _6 \bi _{10} m_1^2+12  \ai _2 \ai _{10} \bi _7 \bi _{10} m_1^2+12  \ai _3 \ai _8 \bi _{10}^2 m_1 m_2+12  \ai _{10}^2 \bi _3 \bi _8 m_1 m_2\nn
+&12 \sqrt{3}  \ai _6 \ai _{10} \bi _3 \bi _8 m_1 m_2+9  \ai _8 \ai _9 \bi _4 \bi _8 m_1 m_2-6 \sqrt{3}  \ai _7 \ai _{10} \bi _4 \bi _8 m_1 m_2+9  \ai _8 \ai _9 \bi _3 \bi _9 m_1 m_2\nn
-&6 \sqrt{3}  \ai _7 \ai _{10} \bi _3 \bi _9 m_1 m_2+9  \ai _4 \ai _8 \bi _8 \bi _9 m_1 m_2+9  \ai _3 \ai _9 \bi _8 \bi _9 m_1 m_2+12 \sqrt{3}  \ai _3 \ai _8 \bi _6 \bi _{10} m_1 m_2\nn
-&6 \sqrt{3}  \ai _4 \ai _8 \bi _7 \bi _{10} m_1 m_2-6 \sqrt{3}  \ai _3 \ai _9 \bi _7 \bi _{10} m_1 m_2+6 \sqrt{3}  \ai _5 \ai _6 \bi _8^2 m_2^2+6  \ai _5 \ai _{10} \bi _8^2 m_2^2+8  \ai _5 \ai _{10} \bi _{10}^2 m_2^2\nn
+&6 \sqrt{3}  \ai _8^2 \bi _5 \bi _6 m_2^2-24  \ai _6 \ai _{10} \bi _5 \bi _6 m_2^2-6 \sqrt{3}  \ai _8 \ai _9 \bi _5 \bi _7 m_2^2+12  \ai _7 \ai _{10} \bi _5 \bi _7 m_2^2-6 \sqrt{3}  \ai _5 \ai _7 \bi _8 \bi _9 m_2^2\nn
+&6  \ai _8^2 \bi _5 \bi _{10}m_2^2+8  \ai _{10}^2 \bi _5 \bi _{10}m_2^2-24  \ai _5 \ai _6 \bi _6 \bi _{10}m_2^2+12  \ai _5 \ai _7 \bi _7 \bi _{10} m_2^2\Big)   ,
\eal
which should hold for any $\ai_i$ and $\bi_i$.

Now, we will see that by choosing some special $\ai_i$ and $\bi_i$ we can constrain $d_1$ and $d_2$ to be zero. If we choose $\ai_1=1$, $\ai_{10}=1$, $\bi_1=1$, $\bi_5=1$ and the other $\ai_i$, $\bi_i$ to be zero, and this above bound reduces to
\beq
\frac{1}{m_2^2 M_1^2}d_1>0   .  
\eeq
Separately, we can choose $\ai_5=1$, $\ai_{10}=1$, $\bi_5=1$ and the other $\ai_i$, $\bi_i$ to be zero, and obtain that
\beq
-\frac{m_1^2+m_2^2}{3 m_1^2 m_2^2 M_1^2}d_1>0  .
\eeq
These two inequalities are written as strict inequalities, which is really only true for the full amplitude. For our leading tree amplitude, we should allow for the posibility of an equality. Thus, combining these two inequalities, we can conclude that
\be
d_1=0  ,
\ee
and the $\epi \epi hhhf$ term is not compatible with the Wilsonian UV completion. Since the theory is formally symmetric in exchanging $h$ and $f$, the $d_2$ term must also vanish
\be
 d_2=0   .
\ee
The theory then reduces to two copies of uncoupled linear Fierz-Pauli Lagrangians. This closes up a loop hole left in the previous results to exclude bi-field pseudo-linear spin-2 theory from positivity bounds.

In summary, the usual elastic positivity bounds constrain bi-field pseudo-linear spin-2 theory from Lagangian (\ref{pseudo3}) to (\ref{leftLpseudo}), while generalized elastic positivity bounds further eliminate the $d_1$ and $d_2$ terms. Therefore, we see that bi-field pseudo-linear spin-2 theory is not compatible with positivity bounds, and does not have a UV completion satisfying the standard axiomatic properties of the S-matrix.

\section{Positivity on interacting massive spin-2 theories}

\label{sec:massspin2}

In the previous section, we have applied generalized elastic positivity bounds to interacting pseudo-linear spin-2 theory, which is a simplified example of interacting massive spin-2 theories that concisely illustrates the effectiveness of generalized elastic bounds. The organizing principle of interacting pseudo-linear spin-2 theory is the linearized diffeomorphism invariance, which is respected by the leading Lovelock terms but broken by the leading dRGT potential terms and the special pseudo-linear terms. Arguably, linearized diffeomorphisms are more akin to the gauge symmetries of matter fields, rather than gravitational fields. In this section, we will apply generalized elastic bounds to full-blown/gravitational interacting massive spin-2 theories whose organizing principle is the full nonlinear diffeomorphism invariance that is broken by the full dRGT potential terms.

The broken diffeomorphism invariance can be restored by introducing the Stueckelberg fields whose transformation laws are patterned after nonlinear diffeomorphisms. This also makes the non-GR modes of massive spin-2 theories more manifest. Indeed, when writing the EFT Lagrangian, it is easier to power-counting the sizes of different terms in the Stueckelberg formulation. Similar to the pseudo-linear case, the reason why the dRGT potential terms are chosen is because we want to look at theories with the highest possible cutoff. These terms are the leading interactions in the Wilsonian effective action that is tuned to have the highest possible cutoff, and the tuning is nature in the technical sense \cite{Alberte:2019lnd}.

For simplicity, in this section we will focus on interacting theories with two dynamical fields in 4D, with generalizations to multiple fields briefly considered in Section \ref{sec:multifield}. In the bi-field case, massive spin-2 theories can be classified into two different kinds \cite{Hinterbichler:2012cn, Alberte:2019lnd}: cycle theory and line theory. They will be discussed in the following separately. The names originate from the distinct topologies one can endow on the interactions among the two dynamical fields and the flat background field in the field space of the three vierbein fields $E^\mu{}_a$, $F^\mu{}_a$ and $I^\mu{}_a=\dd^\mu{}_a$. In terms of the three vierbeins, the bi-field massive spin-2 action we will consider is given by
\bal
\label{Smspin2}
g_{*}^{2} S_{\rm mspin-2}&=\frac{M_{1}^{2}}{8} \int \epsilon_{ABCD} E^{A} \wedge E^{B} \wedge R^{CD}[E]+\frac{M_{2}^{2}}{8} \int \epsilon_{ABCD} F^{A} \wedge F^{B} \wedge R^{CD}[F]
\nn
&~~~~~~~~ + \sum_{a,b,c,d=0}^2 T^{abcd}\epi_{ABCD} H_{(a)}^A\wedge H_{(b)}^B\wedge H_{(c)}^C\wedge H_{(d)}^D +...   ,
\eal
where $\epsilon_{ABCD}$ is the flat space Levi-Civita tensor, $R^{AB}$ is the curvature two form,  $T^{abcd}$ are constant, $H^A_{(0)}\equiv I^A,H^A_{(1)}\equiv E^A,H^A_{(2)}\equiv F^A$ are the vierbein fields, $M_1$ and $M_2$ are the strong couping scales of the helicity-2 modes of $E^A$ and $F^A$ respectively, and $...$ stands for the higher derivative terms. The field space graph can be draw as follows: 1) We denote each of the three vierbteins as a node, and so we have node $I$, node $E$ and node $F$; 2) If there is an interaction term between any two of the three veirbeins, draw a line to connect the two veirbeins; For example, if there is a term like $\epi_{ABCD}I^A\wedge E^B \wedge E^C \wedge E^D$ or $\epi_{ABCD} I^A \wedge I^B \wedge E^C \wedge E^D$, we draw a line connecting node $I$ and node $E$; 3) If there is an interaction term between the three veirbeins such as $\epi_{ABCD}I^A\wedge E^B \wedge E^C \wedge F^D$, we draw a Y shape connection to connect the three nodes. Then a line theory is a graph where we have a line connecting node $I$ and node $E$ (equivalently $F$) and a line connecting node $E$ and node $F$, and a cycle theory is a graph that is not a line.

To turn the vierbein formulation of a massive spin-2 theory into a metric formulation of the theory, we need to impose the symmetric vierbein conditions, which are different for a cycle theory and a line theory, as we show below. For a line theory, the symmetric viertein conditions can be obtained automatically by treating the non-metric, redundant degrees of freedom in the vierbeins as auxiliary fields and integrating them out, similar to that in GR or (single field) dRGT massive gravity. Indeed, for a line theory, because of the simple pairwise connection in the graph, we can effectively separate the action into two sectors of dRGT massive gravity, and thus many properties of dRGT theory naturally follow.  For a cycle theory, on the other hand, its vierbein formulation with and without the symmetric vierbein conditions are in fact not equivalent, that is, a cycle theory with the symmetric veirbein conditions and a cycle theory without the symmetric veirbein conditions are two different theories. We focus on cycle theories with the symmetric veirbein conditions in this paper . It is also worth pointing out that while a line theory is free of the BD ghost, like (single field) dRGT massive gravity, both a cycle theory with and without the symmetric vierbein conditions contain the BD ghost, whose mass is around the scale of the cutoff \cite{Alberte:2019lnd}. Therefore, in the EFT approach, this ghost in a cycle theory is not part of the physical spectrum of the low energy theory and {\it a priori} does not indicate any pathology.

\subsection{Cycle theory}

As just mentioned above, we are considering massive spin-2 theories that have a metric formulation. So the action (\ref{Smspin2}) should be accompanied by appropriate constraints on the vierbeins. We can impose the following symmetric vierbein conditions \cite{Alberte:2019lnd}
\be
\label{cycleveirC}
\eta_{A[B}E^A{}_{\mu]} = 0 ,~~~~\eta_{A[B}F^A{}_{\mu]} = 0   ,
\ee
which means we can introduce symmetric perturbative veirbein field $h_{B\mu}$ and $f_{B\mu}$ as follows
\be
\eta_{AB}E^A{}_{\mu} = \eta_{B\mu} + \f{h_{B\mu}}{M_1},~~~\eta_{AB}F^A{}_{\mu} = \eta_{B\mu} + \f{f_{B\mu}}{M_1}  ,
\ee
satisfying the symmetric conditions $h_{B\mu}=h_{\mu B},f_{B\mu}=f_{\mu B}$. Then the two metrics associated with $E^A{}_\mu$ and $F^A{}_{\mu}$ are given by
\bal
g^{(1)}_{\mu\nu} &= E^A{}_\mu E^B{}_\nu \eta_{AB} = \(\eta_{\mu\ri}+\f{h_{\mu\ri}}{M_1}\)\eta^{\ri\si} \(\eta_{\si\nu}+\f{h_{\si\nu}}{M_1}\)  ,
\\
g^{(2)}_{\mu\nu} &= F^A{}_\mu F^B{}_\nu \eta_{AB}= \(\eta_{\mu\ri}+\f{f_{\mu\ri}}{M_2}\)\eta^{\ri\si} \(\eta_{\si\nu}+\f{f_{\si\nu}}{M_2}\)  .
\eal
We emphasize that this is different from the usual expansion of the metric $g_{\mu\nu}=\eta_{\mu\nu}+h_{\mu\nu}/M$.\;\footnote{To connect to the square root structure of the dRGT potential terms in the metric formulation, we note that
\bal
\epi_{ABCD} H_{(a)}^A\wedge H_{(b)}^B\wedge H_{(c)}^C\wedge H_{(d)}^D &= \epi_{ABCD} \epi^{\mu\nu\ri\si}  H_{(a)\mu}^A H_{(b)\nu}^B H_{(c)\ri}^C H_{(d)\si}^D\d^4 x
\nn
&= \epi_{ABCD} \epi^{\mu\nu\ri\si}   \eta^{AA'}H^{(a)}_{A'\mu}  \eta^{BB'}H^{(b)}_{B'\nu} \eta^{CC'}H^{(c)}_{C'\ri} \eta^{DD'}H^{(d)}_{D'\si}\d^4 x
\nn
&= \epi_{ABCD} \epi^{\mu\nu\ri\si}   \sqrt{\eta^{-1}g^{(a)}}|^A_{\mu}  \sqrt{\eta^{-1}g^{(b)}}|^B_{\nu}  \sqrt{\eta^{-1}g^{(c)}}|^C_{\ri}  \sqrt{\eta^{-1}g^{(d)}}|^D_{\si} \d^4 x  ,
\eal
where here $\epi^{\mu\nu\ri\si}$ is also the flat space Levi-Civita tensor.}.

In the following, since we will be working around the Minkowski background, we shall view the $A,B,C,...$ indices and $\mu,\nu,\ri,...$ indices as of the same type. In terms of the symmetric perturbative veirbein fields $h_{\mu\nu}$ and $f_{\mu\nu}$, the generic cycle theory (\ref{Smspin2}) can be written as \cite{Alberte:2019lnd}
\bal
g_*^2\mc{L}_{\mathrm{cycle}} =~&
\frac{M_{1}^{2}}{2} \sqrt{-g^{(1)}} R(g^{(1)}) +\frac{m_1^2}{4}
\[\epi \epi IIhh+\frac{\ki_3^{(1)}}{M_1}\epi \epi Ihhh+\frac{\ki_4^{(1)}}{M_1^2}\epi \epi hhhh\]
\nn
~+& \frac{M_{2}^{2}}{2} \sqrt{-g^{(2)}} R(g^{(2)}) +\frac{m_2^2}{4}
\[\epi \epi IIff+\frac{\ki_3^{(2)}}{M_2}\epi \epi Ifff+\frac{\ki_4^{(2)}}{M_2^2}\epi \epi ffff\]
\nn
~+&\frac{m_2^2 c_1}{2M_1}\epi \epi Ihhf+\frac{m_2^2 c_2}{2M_2}\epi \epi Iffh+\frac{m_2^2\li}{2M_1 M_2}\epi \epi hhff+\frac{m_2^2 d_1}{4M_1^2}\epi \epi hhhf+\frac{m_2^2 d_2}{4M_2^2}\epi \epi hfff+...  ,
\label{cycle}
\eal
where $\mc{L}_{\mathrm{GR}}(g^{(1,2)},M_{1,2})$ are the Einstein-Hilbert term for metric $g^{(1)}_{\mu\nu}$ and $g^{(2)}_{\mu\nu}$ and other symbols are defined similar to that in the interacting pseudo-linear theory discussed previously (see the explanation below \eref{FPlinear}). However, a major difference here is that in a generic cycle theory while the strong coupling scale between the helicity-0 and helicity-2 modes is still $\Lambda_3$, the interactions from the helicity-0 and helicity-1 modes lower the cutoff of theory to $\Lambda_{7/2}=(m^{5/2}M)^{7/2}$ \cite{Alberte:2019lnd}. The $\Lambda_3$ cutoff is nontheless achievable if we tune interactions between different spin-2 fields, {\it i.e.}, the $c_i,\li,d_i$ coefficients, to be $\mc{O}(m/\Lambda_3)$, on which generalized positivity bounds will be discussed in Section \ref{sec:Li3theory}.

\subsection{Positivity on cycle theory}

Forward elastic positivity bounds from scatterings of $hh\to hh$ and $hf\to hf$ have been considered in \cite{Alberte:2019xfh}. Here we apply generalized elastic positivity bounds, which involve information from inelastic scatterings, to constrain the parameter space of the cycle theory. Similar to the pseudo-linear theory, there is also a formal symmetry in the cycle theory:
\beq
h\leftrightarrow f,~m_1\leftrightarrow m_2,~M_1\leftrightarrow M_2,~\ki_3^{(1)}\leftrightarrow \ki_3^{(2)},~\ki_4^{(1)}\leftrightarrow \ki_4^{(2)}, m_2^2 c_1 \leftrightarrow m_2^2 c_2,~ m_2^2d_1 \leftrightarrow m_2^2 d_2   .
\eeq
Thus, to obtain the generalized elastic positivity bounds, we only need to calculate half of the $16$ amplitudes. Again, we need to regularize the kinematic branch points of the inelastic amplitudes as prescribed in \eref{barM}, and the kinematic poles of the superposed elastic amplitude is regularized by the overall factor
\beq
\Gi(s)=s^2\(s-4m_1^2\)^2\(s-4m_2^2\)^2\[s-(m_1+m_2)^2\]^4\[s-(m_1-m_2)^2\]^4  .
\eeq

\begin{figure}
\centering
\includegraphics[width=0.5\textwidth]{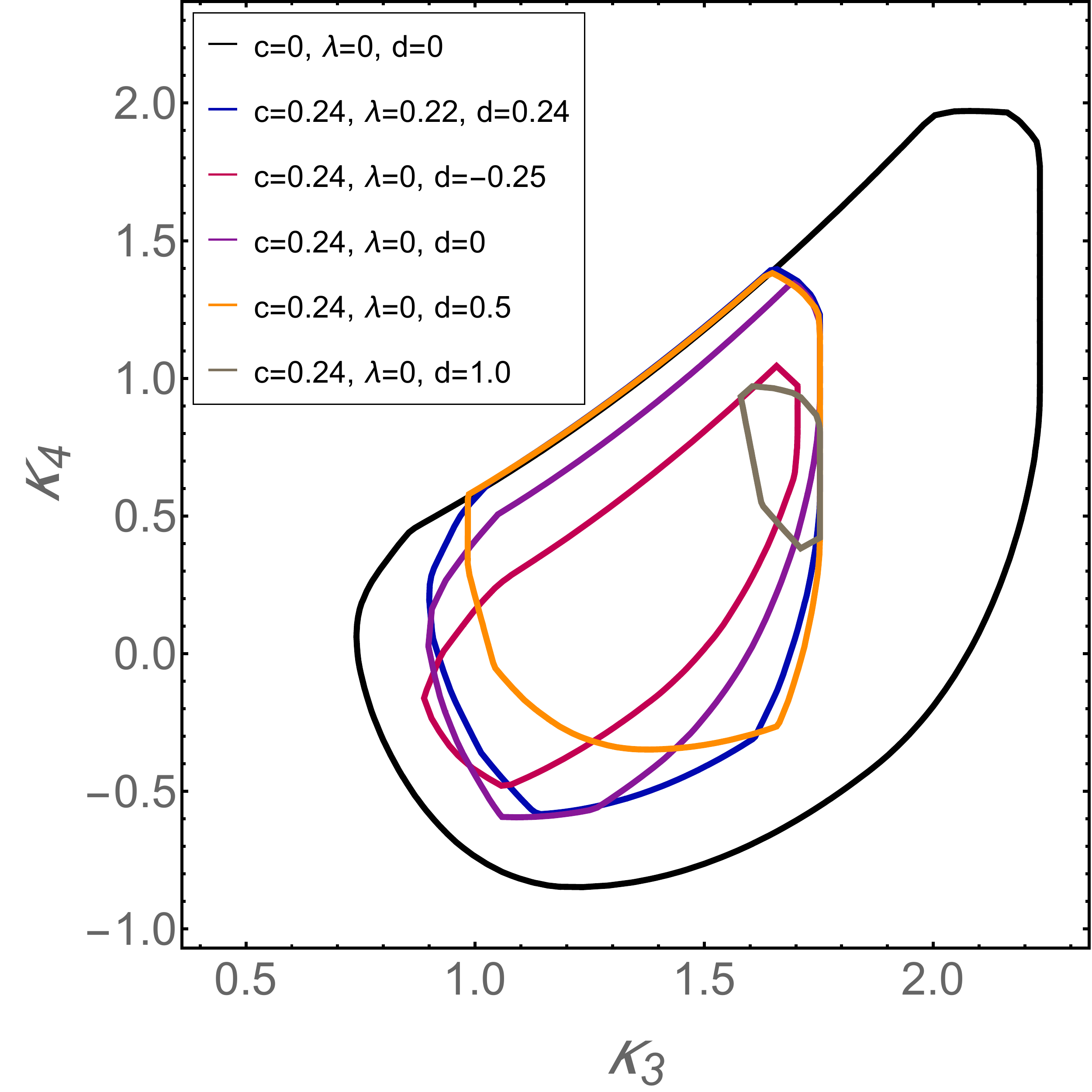}
\caption{Positive regions in the $(\ki_3,\ki_4)$-plane for different $c$, $\li$ and $d$ (the regions within the solid lines are allowed by generalized elastic positivity bounds) in the $\mathbb{Z}_2$ symmetric case. The black line (the largest region) corresponds to the elastic positivity bounds in dRGT gravity. The size of the cross section of this region largely depends on $c$, while $d$ can dramatically change the shape of this region. At $d=d_{min}\simeq -0.5$ or $d=d_{max}\simeq 1.4$, this region shrinks to a point.}
\label{fig:k3k4}
\end{figure}

The first and second line of \eref{cycle} are the terms that appear in dRGT gravity, each of which has $2$ independent Wilson coefficients, $\ki^{(i)}_3$ and $\ki^{(i)}_4$, in additional to the masses, $m_1$ and $m_2$. The third line of \eref{cycle} is the interaction between $h$ and $f$. It is characterized by $5$ independent parameters: $c_1$, $c_2$, $\li$, $d_1$ and $d_2$. Adopting the notation of \cite{Alberte:2019xfh} for an easy comparision, we re-write the parameters as follows
\be
m_1=xm, ~m_2=m, ~M_1=\gi M,~ M_2=M .
\ee
We find that $m$ and $M$ factor out in all the $f_{ijkl}$ defined in \eref{dispersion} as an overall factor of $m^{-2}M^{-2}$. Therefore, we have $11$ dimensionless parameters in total: $x$, $\gi$, $\ki_3^{(1)}$, $\ki_3^{(2)}$, $\ki_4^{(1)}$, $\ki_4^{(2)}$, $c_1$, $c_2$, $\li$, $d_1$ and $d_2$. Note that using elastic positivity bounds Ref.~\cite{Alberte:2019xfh} has put constraints on all these parameters except $d_1$ and $d_2$, since $d_1$ and $d_2$ do not contribute to tree level elastic amplitudes. We will see that these two parameters are constrained in the generalized elastic positivity bounds.

For a given set of Wilson coefficients, the left hand side of the generalized elastic positivity bounds $f_{\ai\bi}>0$ is a quartic polynomial in terms of 20 variables ($\ai_i$ and $\bi_i$). We need to evaluate all possible $\ai_i$ and $\bi_i$ to find the strongest/tightest bounds for this set of Wilson coefficients. Solving quartic inequalities is NP-hard. While for a problem with a less number of variables some analytic methods may be used (see {\it e.g.},~\cite{Bi:2019phv, Yamashita:2020gtt}), numerical methods are often the normal comprise for higher dimensional cases. One naive approach would be use the uniform Monte Carlo method to sample different $\ai_i$ and $\bi_i$ to check whether a given set of Wilson coefficients satisfy all the positivity bounds. As long as the bound is violated for one set of  $\ai_i$ and $\bi_i$, we conclude that this set of Wilson coefficients is in contradiction with positivity. However, we also need to sample the space of the Wilson coefficients, which is also higher dimensional. So this method is inefficient.

\begin{figure}
\centering
\includegraphics[width=0.5\textwidth]{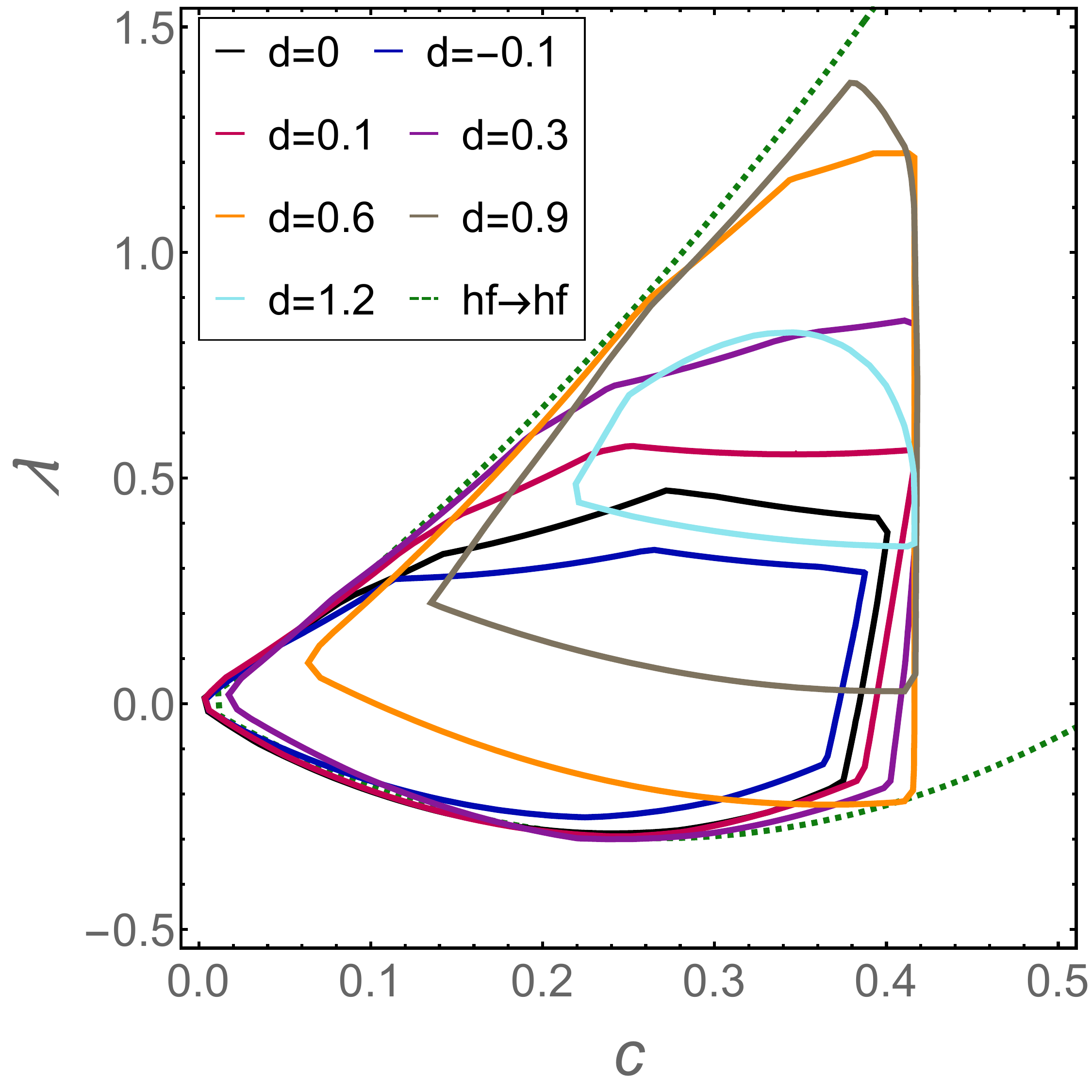}
\caption{Positive regions in the $(c,\li)$-plane for different $d$ with $\ki_3=1.4$, $\ki_4=0.36$ (the regions within the solid lines are allowed by generalized elastic positivity bounds) in the $\mathbb{Z}_2$ symmetric case. The green dashed line is the positivity bound from elastic scattering $hf\rightarrow hf$. We see that the positive cross section changes dramatically with $d$ in this plane.}
\label{fig:cla}
\end{figure}

A better way to check whether the bound $f_{\ai\bi}>0$ is satisfied for a given set of Wilson coefficients is to formulate the problem in terms of an autonomous dynamical system \cite{Cheung:2016yqr}. Let $X_I=(\ai_1,\ai_2,...,\ai_{10},\bi_1,\bi_2,...,\bi_{10})$ depend on some fictitious time $t$ and evolve according the following set of ordinary differential equations:
\be
\ddt X_I = -\f{\pd f_{\ai\bi}}{\pd X_I}   .
\ee
Randomly choosing a set of $X_I$ in the interval of $[-1,1]$ as the initial condition, we evolve this dynamical system for a sufficient time interval. The evolution will generically lead $X_I$ to values that make $f_{\ai\bi}$ smaller, since
\be
\ddt f_{\ai\bi} = \sum_I \f{\pd f_{\ai\bi}}{\pd X_I} \f{\d X_I}{\d t} = - \sum_I \f{\pd f_{\ai\bi}}{\pd X_I}\f{\pd f_{\ai\bi}}{\pd X_I}<0   .
\ee
In cases where $f_{\ai\bi}$ can be negative, this is method is usually very efficient to find a $X_I$ that does so.
Complications may arise if there are multiple equilibria (defined by points $X_I$ satisfying ${\pd f_{\ai\bi}}/{\pd X_I}=0$) for this dynamical system. $X_I=0$ is an equilibrium of this system, and sometimes there may be other equilibria. Actually, since $f_{\ai\bi}$ is a homogenous function of $X_I$, if $X_I=c_I$ is an equilibrium, then the line connecting $X_I=0$ and $X_I=c_I$ is a continuous set of equilibria. Also, this method may fail to find negative $f_{\ai\bi}$ if the initial $X_I$ happens to be on a trajectory that passes very close by the $X_I=0$ equilibrium, near which the evolution is very slow due to the very property of an equilibrium. So it is prudent to try a few set of $X_I$ as the initial conditions.

As we see above, the total parameter space even in the bi-field massive spin-2 case is very large. For simplicity, we will focus on models with a reduced number of parameters, that is, we apply the generalized elastic positivity bounds to the case of a $\mathbb{Z}_2$ symmetry, a $\mathbb{Z}_2$ symmetry but with $x\neq 1$ and a $\mathbb{Z}_2$ symmetry but with $\gi\neq 1$.

\begin{figure}
\centering
\includegraphics[width=0.4\textwidth]{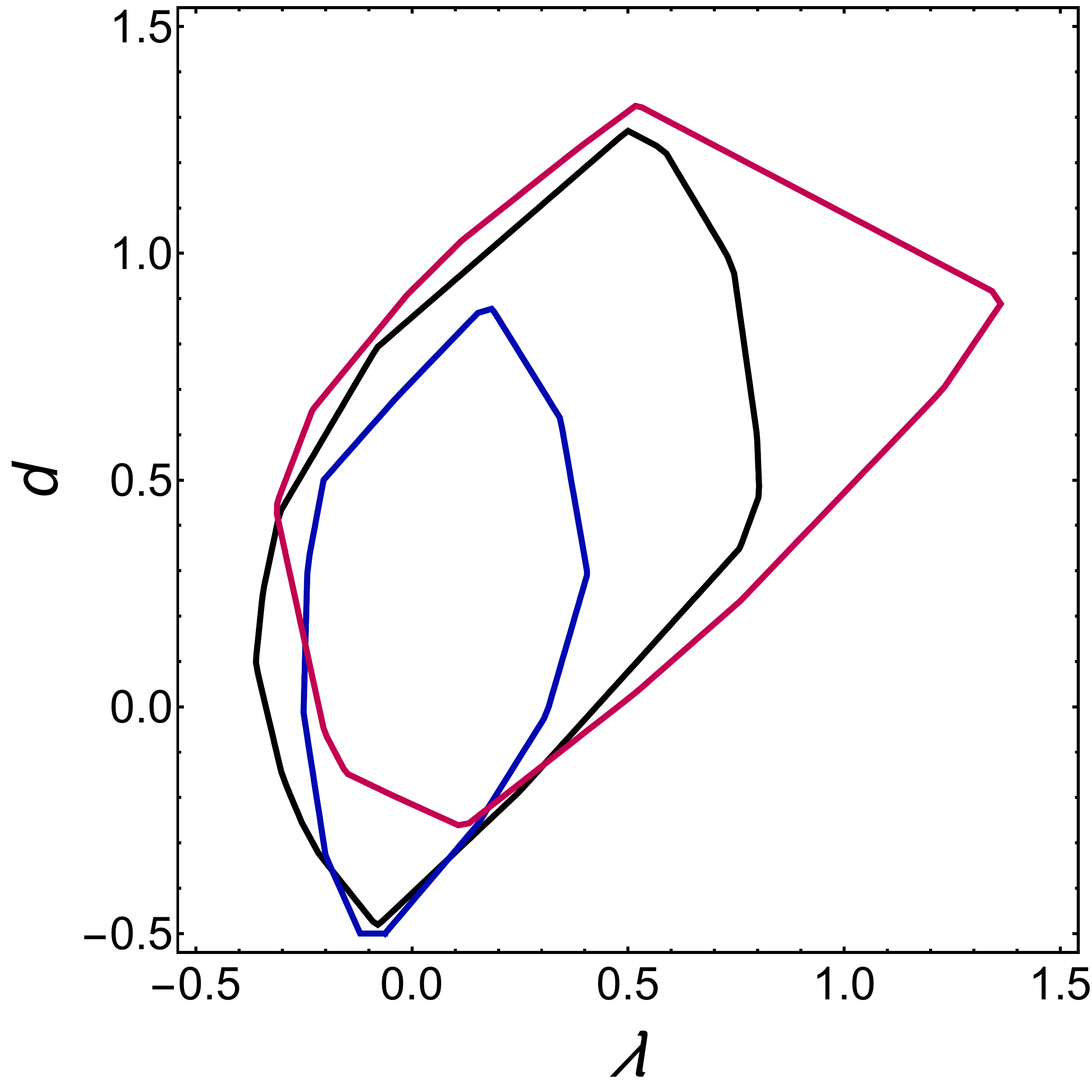}
\includegraphics[width=0.4\textwidth]{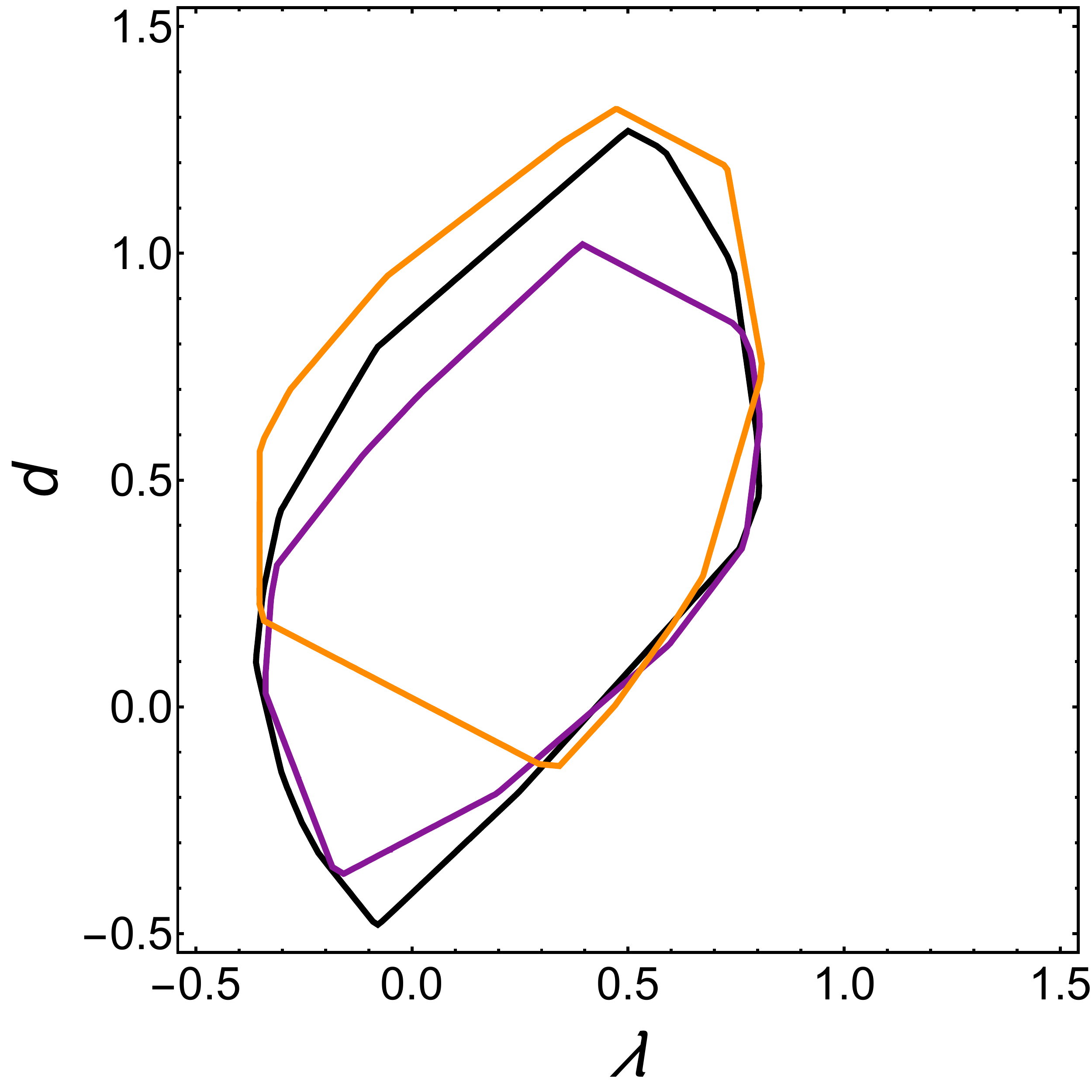}
\\
\includegraphics[width=0.4\textwidth]{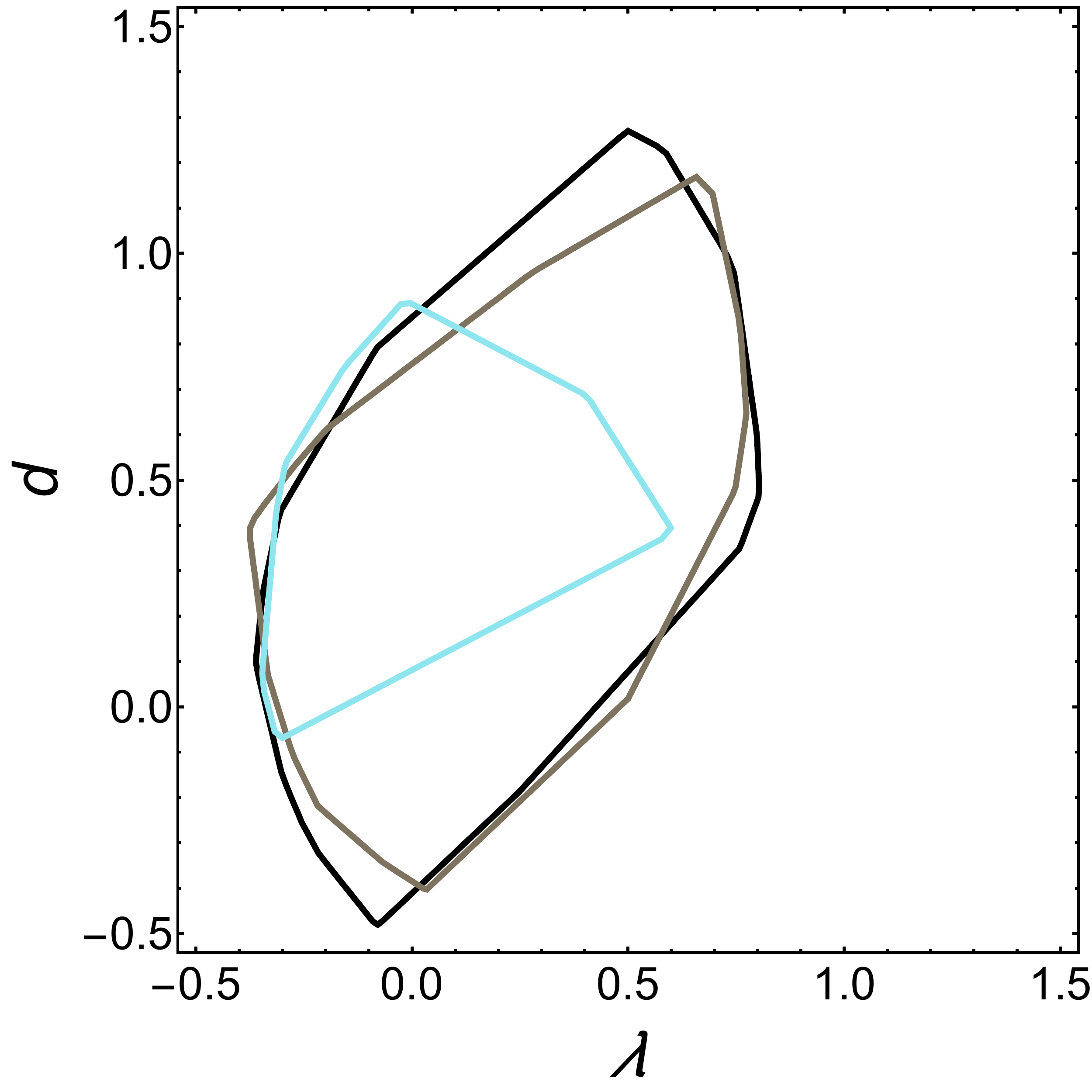}
\includegraphics[width=0.4\textwidth]{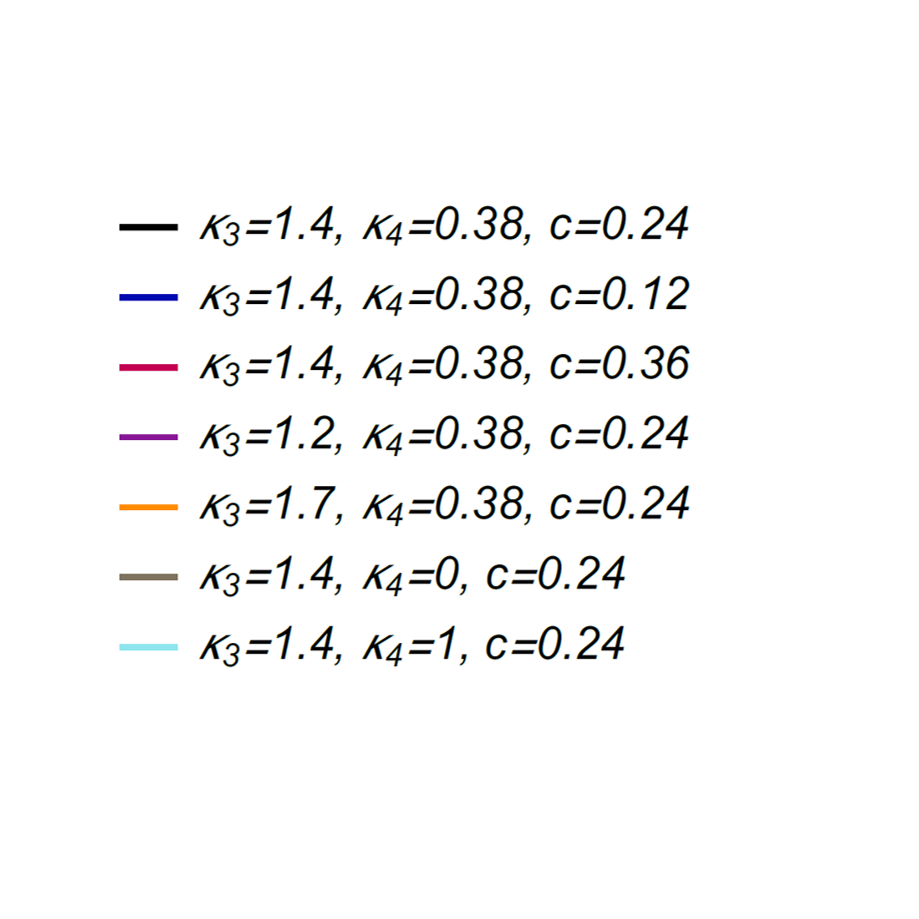}
\caption{Positive regions in the $(\li,d)$-plane with different $\ki_3$, $\ki_4$ and $c$ (the regions within the solid lines are allowed by generalized elastic positivity bounds) in the $\mathbb{Z}_2$ symmetric case. The black line is when parameters other than $\li$ and $d$ take their approximate central values in \eref{centralV}. The top left, top right and bottom plot show the influence of changing $\ki_3$, $\ki_4$, and $c$ respectively.}
\label{fig:lad}
\end{figure}

~\\
\noindent{\bf $\bullet~\mathbb{Z}_2$ symmetric case}

The $\mathbb{Z}_2$ symmetric theory we consider here is the one where we identify the following parameters
\be
x=1, ~\gi=1, ~\ki_3^{(1)}=\ki_3^{(2)}=\ki_3, ~\ki_4^{(1)}=\ki_4^{(2)}=\ki_4,~ c_1=c_2=c,~d_1=d_2=d .
\ee
In other words, the theory is invariant under exchanging $h\leftrightarrow f$ in the $\mathbb{Z}_2$ symmetric case, and we are now left with $5$ parameters: $\ki_3$, $\ki_4$, $c$, $\li$, $d$.

We shall find the allowed positive region numerically with the dynamical system method mentioned above.
Using the generalized elastic bounds, we find that the allowed region in the 5D parameter space is simply connected. The approximate center of this connected region is
\beq
\label{centralV}
\bar{\ki}_3=1.4,~\bar{\ki}_4=0.34,~\bar{c}=0.24,~\bar{\li}=0.23,~\bar{d}=0.36  ,
\eeq
and the approximate maximum and minimum values of these parameters are
\beq
0.8\lesssim\ki_3\lesssim2.2,~
-0.8\lesssim\ki_4\lesssim1.7,~
0\lesssim c\lesssim0.6,~
-0.4\lesssim\li\lesssim1.7,~
-0.5\lesssim d\lesssim1.4   .
\eeq
To estimate the volume of this 5D positive region, we choose a cuboid $V$ encircling the above region and use the simple Monte Carlo method to sample this $V$. If the total number of the sampling points is $N_T$ and the number of the points falling into the positive region is $N_+$, then the volume of the positive region is
\be
\label{Omega5}
\Omega^{(5)} = \f{N_+}{N_T}V\simeq 0.3 
\ee
whose value is independent of the choice of $V$ if the sampling is sufficient good. Of course, to get better accuracy, $V$ should not be much greater than the cuboid specified by the approximate maximum and minimum values above.

\begin{figure}
\centering
\includegraphics[width=0.4\textwidth]{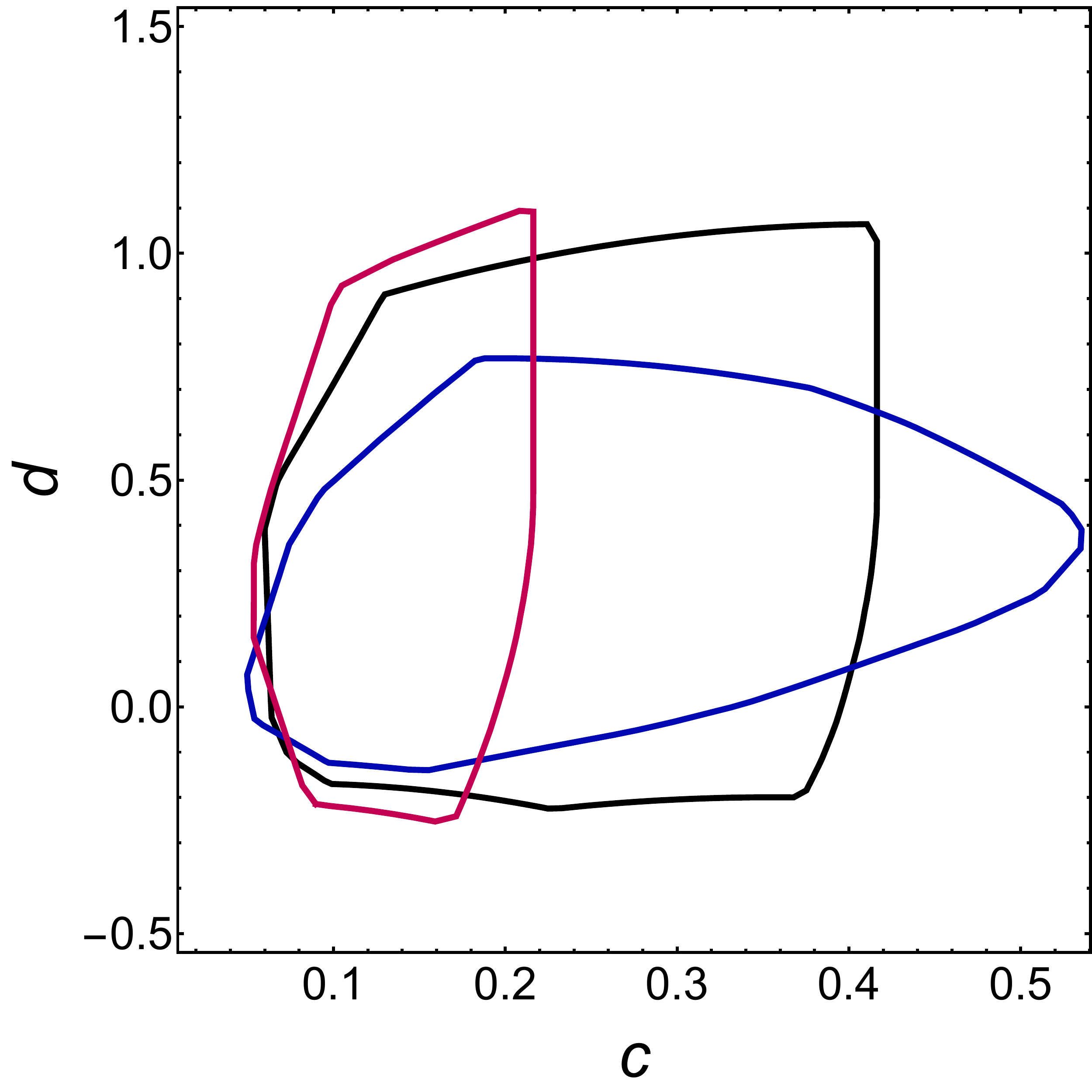}
\includegraphics[width=0.4\textwidth]{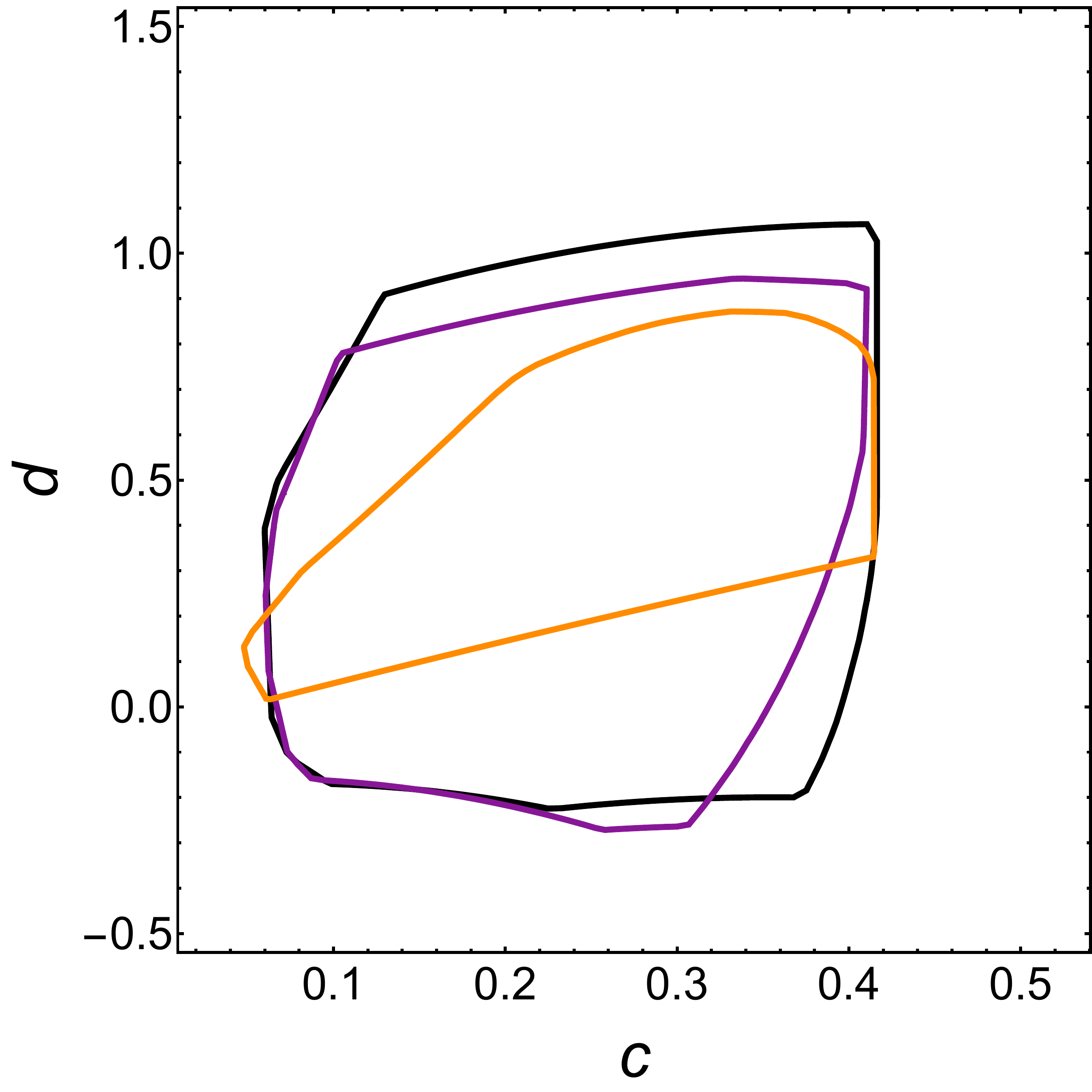}
\\
\includegraphics[width=0.4\textwidth]{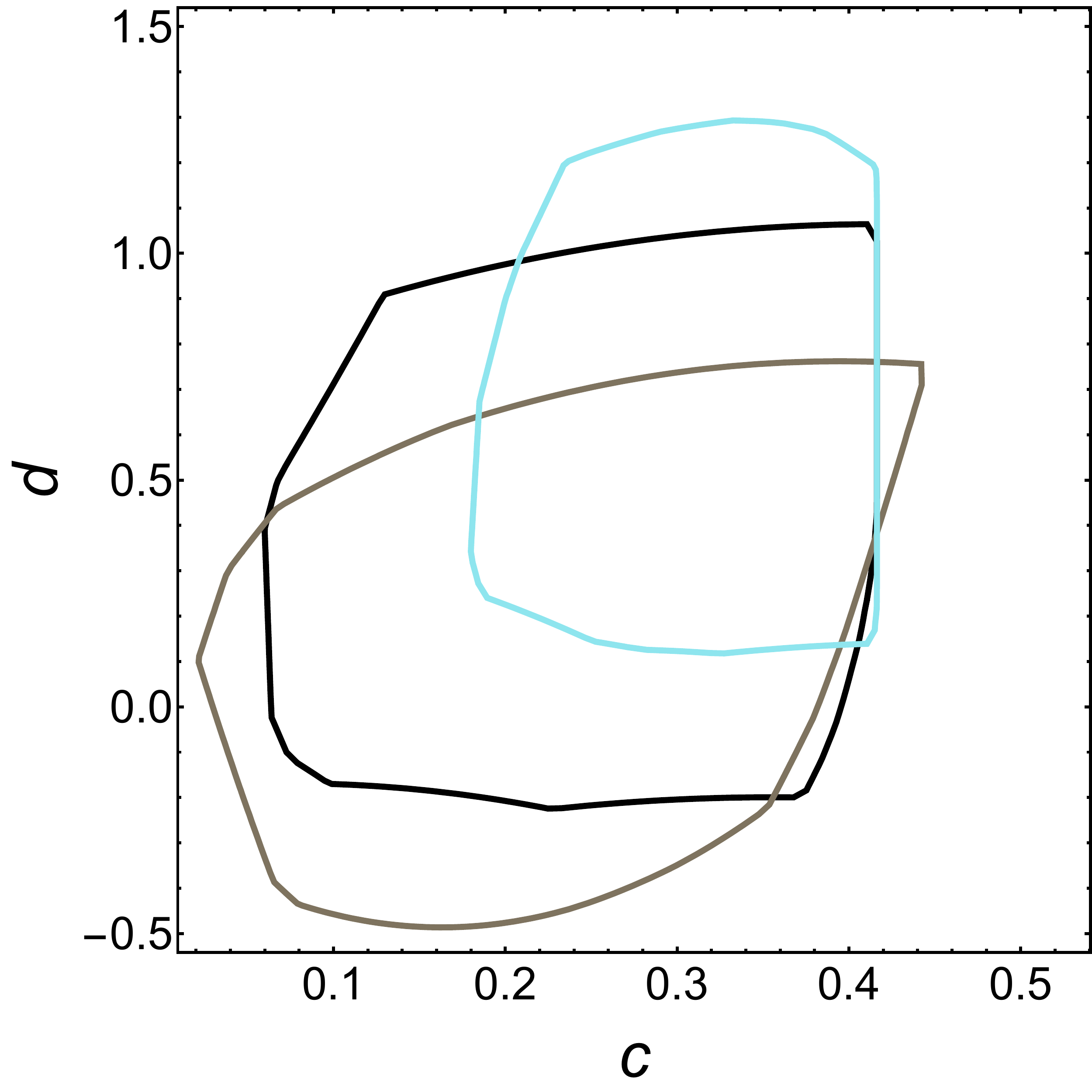}
\includegraphics[width=0.4\textwidth]{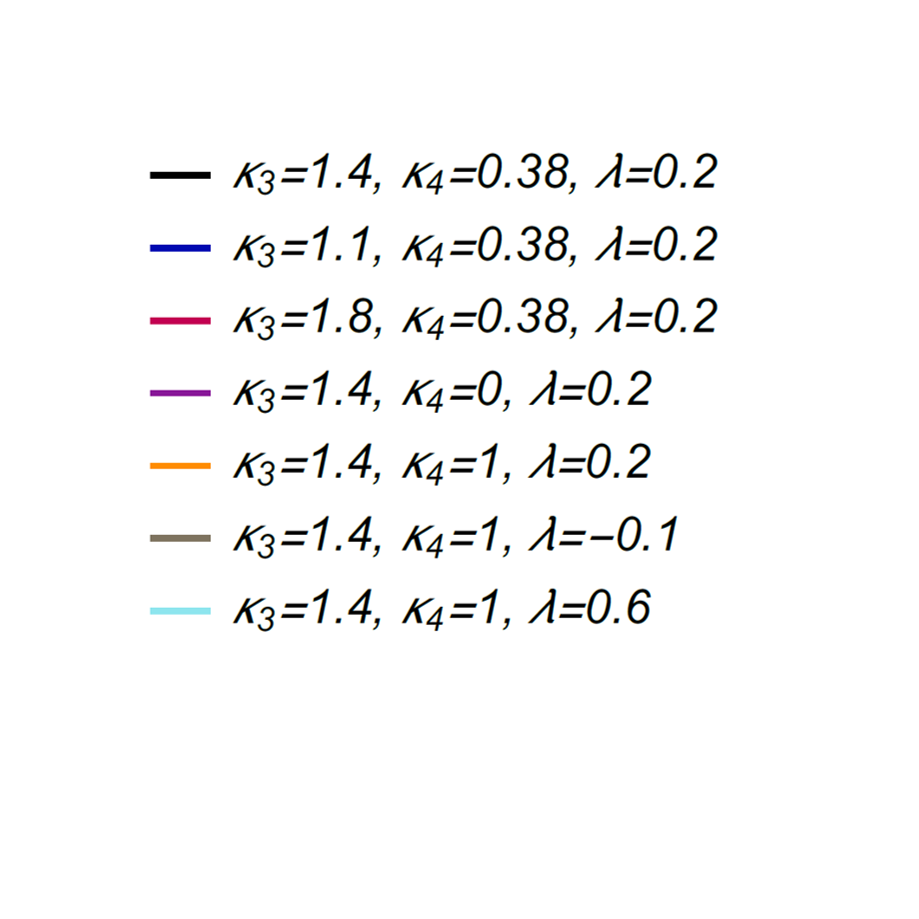}
\caption{Positive regions in the $(c,d)$-plane with different $\ki_3$, $\ki_4$ and $\li$ (the regions within the solid lines are allowed by generalized elastic positivity bounds) in the $\mathbb{Z}_2$ symmetric case. The black line is when parameters other than $c$ and $d$ take their approximate central values in \eref{centralV}. The top left, top right and bottom plot show the influence of changing $\ki_3$, $\ki_4$, and $\li$ respectively.}
\label{fig:cd}
\end{figure}

For a comparison, in \cite{Alberte:2019xfh}, $\ki_3$, $\ki_4$, $c$ and $\li$ are constrained to a finite region, but $d$ is unconstrained, and the usual elastic bounds of $hh\rightarrow hh$, $ff\rightarrow ff$ and $hf\rightarrow hf$ restrict the $4$-D parameters space to a region with $\Omega^{(4)}\approx2.6$. With our generalized bounds, $d$ is now constrained to $-0.5\lesssim d\lesssim1.4$. If we were to restrict $d$ to be in this range and compare the corresponding 5D volume to the $\Omega^{(5)}$ from the generalized bounds, we would find that 
\be
\f{\Omega^{(5)}}{\Omega^{(4)}(d_{\rm max}-d_{\rm min})} \simeq 6\%  .
\ee
Therefore, we see that the new bounds do not only constrain $d$ from a free parameter to be within an interval, and the allowed volume for the other 4 parameters is also greatly reduced.

We also want to visualize 2D sections of this allowed positive region.  In Fig.~\ref{fig:k3k4}, we show the  allowed positive regions in the ($\ki_3$,$\ki_4$)-plane for different $c$, $\li$ and $d$. The general trend is that the size of the positive cross section largely depends on $c$, while $d$ can dramatically change its shape. This positive region shrinks to a point at $d=d_{min}\simeq -0.5$ or $d=d_{max}\simeq 1.4$. In Fig.~\ref{fig:cla}, we show that the allowed positive regions in the $(c,\li)$-plane for different $d$. The green dashed line is the positivity bound from elastic scattering $hf\rightarrow hf$. We see that as expected by choosing different $d$, our generalized positivity bounds give rise to stronger constraints. For smaller $d$, the allowed region includes the point $c=0, \li=0$, but when $d$ becomes greater, this point is ruled out by positivity, which means the operators corresponding to the $c$ and $\li$ coefficients have to be included in the EFT to have a standard UV completion. In the $(c,\li)$-plane, the allowed positive region changes significantly with $d$. In Fig.~\ref{fig:lad}, we show how the positive region changes by changing $\ki_3$, $\ki_4$ and $c$ separately in the $(\li,d)$ plane, while in Fig.~\ref{fig:cd}, we show how the positive region changes by changing $\ki_3$, $\ki_4$, and $\li$ separately in the $(c,d)$ plane.

\begin{figure}
\centering
\includegraphics[width=0.5\textwidth]{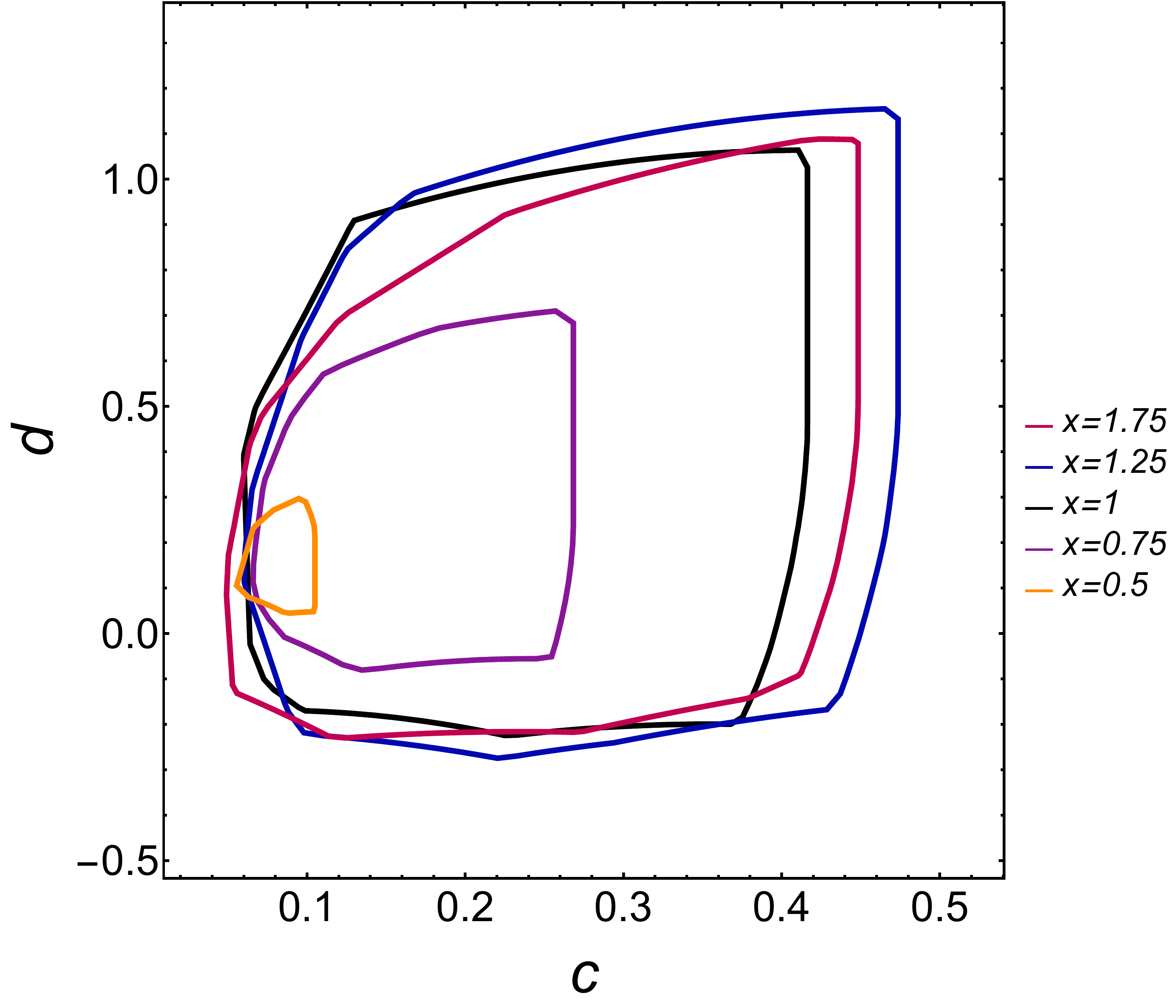}
\caption{Positive regions in the ($c,d$)-plane with different $x$ (the regions within the solid lines are allowed by generalized elastic positivity bounds) in the $\mathbb{Z}_2$ symmetric but $x\neq 1$ case. We choose $\kappa_3=1.4, \kappa_4=0.38, \lambda=0.2$ and $1/2<x<2$.}
\label{fig:cdx}
\end{figure}

\begin{figure}
\centering
\includegraphics[width=0.5\textwidth]{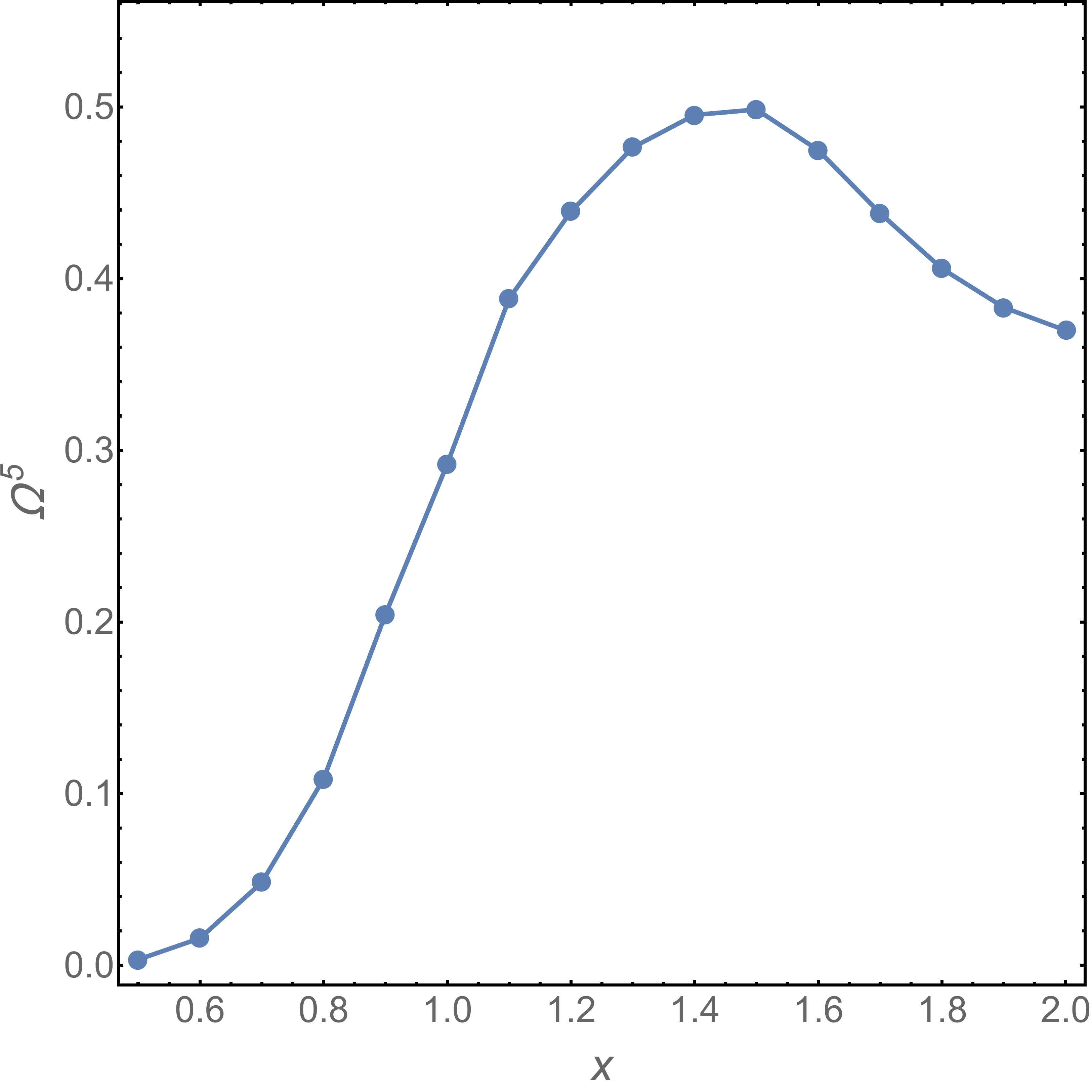}
\caption{Volume of 5D positive region $\Omega^{(5)}$ with different $x$ for $\gi=1$. $\Omega^{(5)}$ is greater when $x$ is close to $1$.}
\label{fig:Oix}
\end{figure}

~\\
\noindent{\bf $\bullet~\mathbb{Z}_2$ symmetric but with $x \neq 1$}:

In Fig.~\ref{fig:cdx}, we show how the positive region changes with $x$. Within the range of $1/2<x<2$, the positive region is smaller in both ends and greater in the middle, while the shape of the positive region remains largely unchanged for different $x$. Again, we can estimate the volume of the positive region in the 5D parameter space of $\ki_3$, $\ki_4$, $c$, $\li$ and $d$, as done in \eref{Omega5} for the $\mathbb{Z}_2$ case, and see how it changes with the value of $x$, which is shown in Fig.~\ref{fig:Oix}. We see that the volume of the 5D region allowed by positivity is again smaller in both ends and greater in the middle.

~\\
\noindent{\bf $\bullet~\mathbb{Z}_2$ symmetric but with $\gi \neq 1$}

\begin{figure}
\centering
\includegraphics[width=0.5\textwidth]{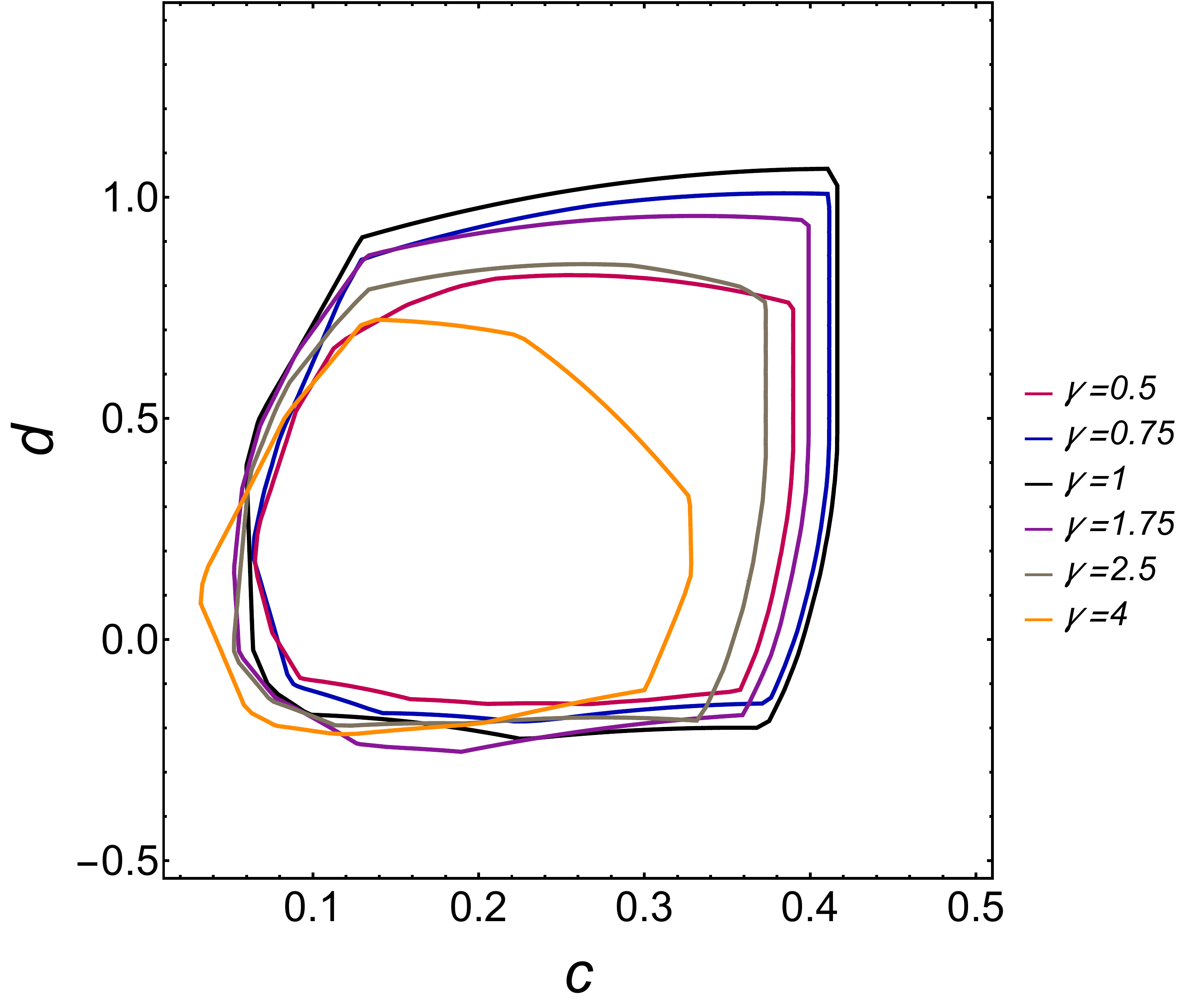}
\caption{Positive regions in the ($c,d$)-plane with different $\gi$ (the regions within the solid lines are allowed by generalized elastic positivity bounds) in the $\mathbb{Z}_2$ symmetric but $\gi\neq 1$ case. We choose $\kappa_3=1.4, \kappa_4=0.38, \lambda=0.2$.}
\label{fig:cdg}
\end{figure}

\begin{figure}
\centering
\includegraphics[width=0.5\textwidth]{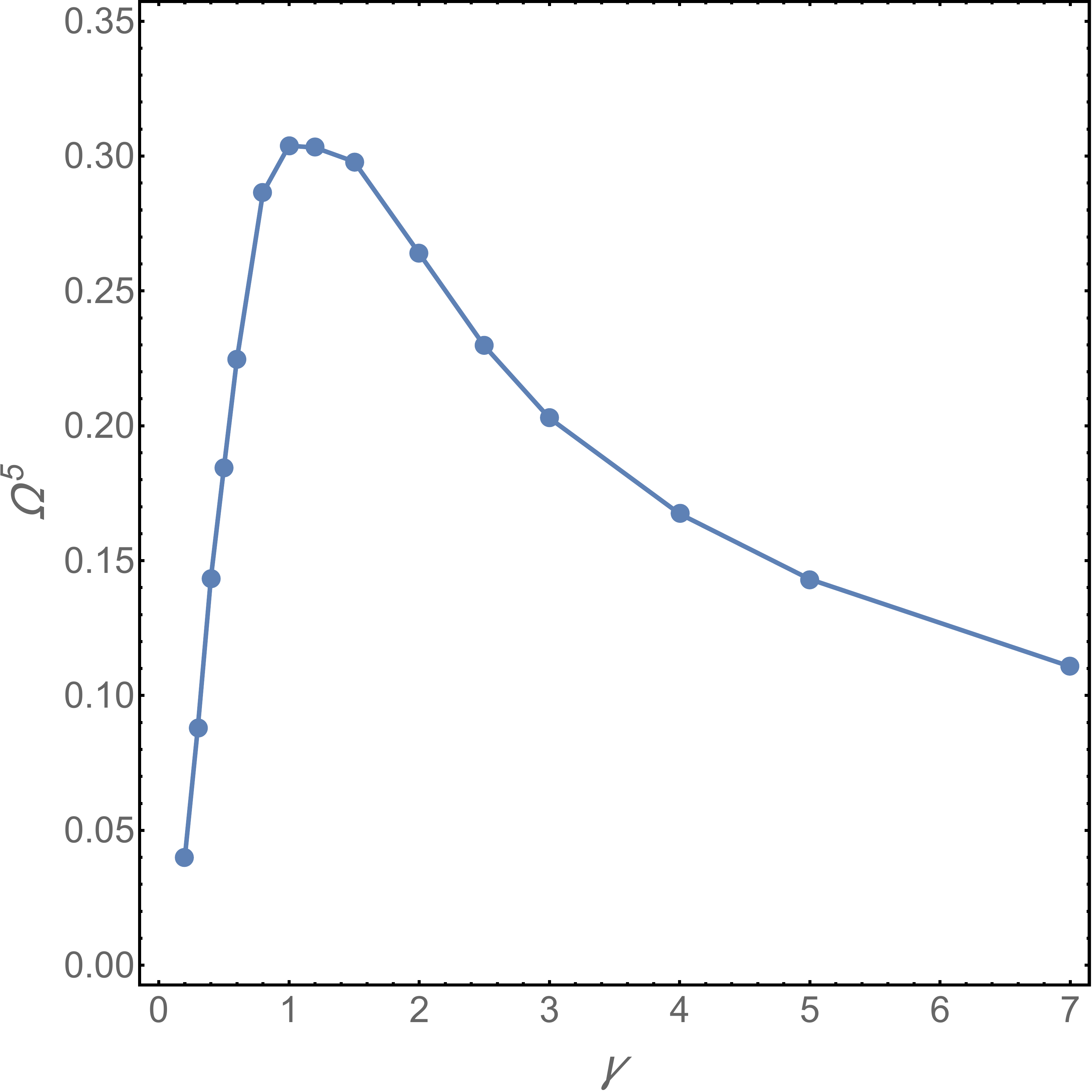}
\caption{Volume of 5D positive region $\Omega^{(5)}$ with different $\gi$ for $x=1$. The maximum of $\Omega^{(5)}$ is at $\gi=1$.}
\label{fig:Oig}
\end{figure}

In Fig.~\ref{fig:cdg}, we show how the positive region changes with $\gi$. Within the range of $1/2<\gi<4$, the positive region is again smaller in both ends and greater in the middle.  We also want to estimate the volume of the positive region in the 5D parameter space of $\ki_3$, $\ki_4$, $c$, $\li$ and $d$, as done in \eref{Omega5} for the $\mathbb{Z}_2$ case, and see how it changes with the value of $\gi$, which is shown in Fig.~\ref{fig:Oig}. We see that the volume of the 5D region allowed by positivity is smaller in both ends and greater in the middle.

\subsubsection{$\Li_3$ cycle theory}

\label{sec:Li3theory}

In the above, we have applied the generalized positivity bounds to a generic cycle theory that has a cutoff at $\Lambda_{7/2}$. By choosing the interactions between $h$ and $f$ to be suppressed by an extra factor of $m/\Lambda_3$, we get a $\Li_3$ cycle theory. That is, in a $\Li_3$ cycle theory, the following parameters in \eref{cycle} are scaled as follows
\beq
\label{L3supp}
c_1= \tilde{c}_1\frac{m}{\Li_3},~ c_2= \tilde{c}_2\frac{m}{\Li_3},~ \li= \tilde{\li}\frac{m}{\Li_3},~ d_1= \tilde{d}_1\frac{m}{\Li_3},~ d_2= \tilde{d}_2\frac{m}{\Li_3}   ,
\eeq
where $\tilde{c}_{1,2},~\tilde{d}_{1,2},~\tilde{\li}\sim \mathcal{O}(1)$. This means that $c_{1,2},~d_{1,2},~\li \ll 1$ but $\ki_{3,4}^{(1,2)}\sim \mathcal{O}(1)$. Because of this fact, for the bi-field case, generalized elastic positivity bounds do not lead to extra new constraints on the theory, compared to the bounds in \cite{Alberte:2019lnd}. Put it another way, we may still use the positivity bounds to first constrain un-tilded coefficients, exactly as we did for the $\Lambda_{7/2}$ case; To obtain the constraints on the tilded coefficients, we need to enlarge the un-tilded coefficients by a large factor $\Lambda_3/m$, which essentially means that there is no effective constraints along these directions, and this renders generalized elastic bounds ineffective to improve the elastic bounds.

To see this, first note that the tuning \eref{L3supp} implies that the positivity bounds from the $hh\rightarrow hh$ (or $ff\rightarrow ff$) amplitudes merely constrain $h$ (or $f$) self-interaction coefficients to leading order, since the contributions to $f_{ijkl}$ from the $c_1^2$, $c_2^2$ terms are suppressed and the leading contributions come from single-field dRGT self interactions. The key fact is that for any definite helicity scattering we have $f_{ijij}\neq 0$ ($i$ and $j$ only representing helicities for a single field) at leading order. On the other hand, for other scattering amplitudes, elastic $hf\rightarrow hf$(or $fh\rightarrow fh$) and inelastic, the leading contributions to  $f_{ijkl}$ come from the $c_{1,2}$, $d_{1,2}$ and $\li$ interactions, which are subleading compared to those of $hh\rightarrow hh$ (or $ff\rightarrow ff$). Therefore, if generalized elastic positivity bounds were to give extra constraints, they should not contain the contributions from $hh\rightarrow hh$ (or $ff\rightarrow ff$) and should come from the bounds
\bal
\left.f_{\ai\bi}\right|_{\ai_6\rightarrow0,\ai_7\rightarrow0,\ai_8\rightarrow0,\ai_9\rightarrow0,\ai_{10}\rightarrow0,\bi_1\rightarrow0,\bi_2\rightarrow0,\bi_3\rightarrow0,\bi_4\rightarrow0,\bi_5\rightarrow0}>0\nn
\({\rm or} \left.f_{\ai\bi}\right|_{\ai_1\rightarrow0,\ai_2\rightarrow0,\ai_3\rightarrow0,\ai_4\rightarrow0,\ai_5\rightarrow0,\bi_6\rightarrow0,\bi_7\rightarrow0,\bi_8\rightarrow0,\bi_9\rightarrow0,\bi_{10}\rightarrow0}>0\)   .
\eal
But these are just the usual elastic bounds from $hf\rightarrow hf$(or $fh\rightarrow fh$). Therefore, for the $\Lambda_3$ cycle theory, the generalized elastic bounds do not give rise to more constraints than those in \cite{Alberte:2019lnd}. However, as we will discuss later, for multiple fields, the generalized elastic bounds can give rise to extra constraints.

\subsection{Line theory}

For the line theory, the vierbein formulation is equivalent to the corresponding metric formulation, which can be obtained by integrating out the redundant local Lorentz fields, that is, imposing the following symmetric vierbein conditions \cite{Alberte:2019lnd}
\be
\eta_{A[B}E^A{}_{\mu]} = 0 ,~~~~\eta_{AB} E^B{}_{[\nu}F^A{}_{\mu]} = 0  .
\ee
 In the metric formulation, the line theory can be written as
\bal
g_*^2\mc{L}_{\mathrm{line}}  &=
\frac{M_{1}^{2}}{2} \sqrt{-g^{(1)}} R(g^{(1)})+\frac{\tilde{m}_{1}^{2} M_{1}^{2}}{4} \sqrt{-\eta} \( \epi\epi I^2 \mc{K}_1^2 + \tilde\ai_3  \epi\epi I \mc{K}_1^3    + \tilde\ai_4  \epi\epi  \mc{K}_1^4\)
 \nn
&+\frac{M_{2}^{2}}{2} \sqrt{-g^{(2)}} R(g^{(2)})+\frac{\tilde{m}_{2}^{2} M_2^{2}}{4} \sqrt{-g^{(1)}} \( \epi\epi I^2 \mc{K}_2^2 + \tilde\bi_3  \epi\epi I \mc{K}_2^3   + \tilde\bi_4  \epi\epi  \mc{K}_2^4 \) +\cdots   ,
\eal
where $\mc{K}_1{}^\mu{}_\nu=\dd^\mu{}_\nu - \sqrt{\eta^{-1}g^{(1)}}\big|^\mu{}_\nu$, $\mc{K}_2{}^\mu{}_\nu=\dd^\mu{}_\nu - \sqrt{g_{(1)}^{-1}g^{(2)}}\big|^\mu{}_\nu $, we have again used the double Levi-Civita contraction defined above ({\it e.g,} $\epi\epi I^2 \mc{K}_1^2= \epi\epi II \mc{K}_1\mc{K}_1$), and $...$ stands for subleading higher derivative terms. Without the higher derivative terms, the line theory is free of the BD ghost to all scales, so in this sense the line theory is the {\it bona fide} multi-field generalization of (single field) dRGT massive gravity, and it is indeed a $\Lambda_3$ theory around the flat background.

Although the symmetric vierbein conditions are now different from those of the cycle theory \eref{cycleveirC}, we can still parametrize the metrics as follows
\bal
g^{(1)}_{\mu\nu} & = \(\eta_{\mu\ri}+\f{\tilde h_{\mu\ri}}{M_1}\)\eta^{\ri\si} \(\eta_{\si\nu}+\f{\tilde h_{\si\nu}}{M_2}\)  ,
\\
g^{(2)}_{\mu\nu} &= \(\eta_{\mu\ri}+\f{\tilde f_{\mu\ri}}{M_1}\)\eta^{\ri\si} \(\eta_{\si\nu}+\f{\tilde f_{\si\nu}}{M_2}\) ,
\eal
with $\tilde h_{\mu\nu},\tilde f_{\mu\nu}$ symmetric, which amounts to field redefinitions for the metrics $\tilde h_{\mu\nu} = M_1 (\eta_{\mu\ri}\mc{K}_1{}^\ri{}_\nu -\eta_{\mu\nu})$ and $\tilde f_{\mu\nu} = M_1 (\eta_{\mu\ri}\mc{K}_2{}^\ri{}_\nu -\eta_{\mu\nu})$. With these redefinitions, while $\mc{K}_1$ simply becomes $\mc{K}_1{}^\mu{}_\nu=\tilde h^\mu{}_\nu/M_1$, $\mc{K}_1$ becomes $\mc{K}_2{}^\mu{}_\nu=\tilde h^\mu{}_\nu/M_1-\tilde f^\mu{}_\nu/M_2+...$
So in this formulation the line theory has a kinematic mixing for the mass terms
\bal
g_*^2\mc{L}_{\mathrm{line}} &\supset
-\frac{1}{2} \tilde{m}_{1}^{2}\left([\tilde{h}^{2}]-[\tilde{h}]^{2}\right) -\frac{1}{2} \f{\tilde{m}_{2}^{2}}{M_1^2+M_2^2}\([(M_1\tilde{f}-M_2\tilde{h})^{2}]-[M_1\tilde{f}-M_2\tilde{h}]^{2}\)  ,
\eal
where the trace $[~ ]$ is defined with the Minkowski metric, {\it e.g.}, $[\tilde h\tilde f]\equiv = \tilde h_\mu{}^\nu \tilde f_\nu{}^\mu=  \tilde h_{\mu\nu} \ei^{\nu\si}\tilde f_{\si\ri}\ei^{\rho\mu}$. Thus, $\tilde h_{\mu\nu}$ and $\tilde f_{\mu\nu}$ are not the mass eigenstates of the theory. To get the mass eigenstates, we diagonalize the perturbative metrics
\be
\begin{pmatrix}
h_{\mu\nu}\\
f_{\mu\nu}
\end{pmatrix}
=
\begin{pmatrix}
\cos\thi &-\sin\thi\\
\sin\thi & \cos\thi
\end{pmatrix}
\begin{pmatrix}
\tilde h_{\mu\nu}\\\tilde f_{\mu\nu}
\end{pmatrix}   ,
\ee
where the mixing angle is given by
\beq
\thi=\f12 \arctan \frac{2 M_1 M_2 \tilde{m}_2^2}{M_1^2(\tilde{m}_1^2-\tilde{m}_2^2)+M_2^2(\tilde{m}_1^2+\tilde{m}_2^2)}  ,
\eeq
The physical masses are given by
\beq
m_{1,2}^2=\frac{1}{2}\(
\tilde{m}_1^2+\tilde{m}_2^2\pm\frac{M_1^2(\tilde{m}_1^2-\tilde{m}_2^2)^2+M_2^2(\tilde{m}_1^2+\tilde{m}_2^2)^2}{M_1^2(\tilde{m}_1^2-\tilde{m}_2^2)+M_2^2(\tilde{m}_1^2+\tilde{m}_2^2)}\mathrm{cos}~2\thi  ,
\)
\label{line}
\eeq
where the $+$ sign is for $m_1$ and the $-$ sign is for $m_2$. After this, the line spin-2 theory for two fields up to quartic order is given by \cite{Alberte:2019xfh}
\bal
\label{linehf}
g_*^2\mc{L}_{\mathrm{line}}
&= \mc{L}_{\mathrm{FP}}(h,m_1)+\mc{L}_{\mathrm{FP}}(f,m_2)+\mc{L}_\mathrm{GR}^{3,4}(\mathrm{cos}\thi h+\mathrm{sin} \thi f,M_1) +\mc{L}_\mathrm{GR}^{3,4}(\mathrm{cos}\thi f-\mathrm{sin} \thi h,M_2)\\
&+\frac{\tilde{m}^2_2}{4M_1}\sum_{n=0}^{3}\ki_{n}^{(3)}\epi \epi I h^{3-n}f^{n}
+\frac{\tilde{m}^2_2}{4M_1^2}\sum_{n=0}^{4}\ki_{n}^{(4)}\epi \epi I h^{4-n}f^{n}
+\frac{\tilde{m}^2_2}{4(M_1^2+M_2^2)}\([(fh)^2]-[f^2 h^2]\)+...  ,
\eal
where $\mc{L}_{\mathrm{FP}}(h,m)$ is the Fierz-Pauli term for field $h_{\mu\nu}$ with mass $m$ and $\mc{L}_\mathrm{GR}^{3,4}(h,M)$ is the standard GR three and four point interactions for field $h_{\mu\nu}$ with cutoff $M$. $\ki_n^{(3,4)}$ in \eref{line} can be written in terms of $\tilde{\ai}_{3,4}$ and $\tilde{\bi}_{3,4}$ and other parameters (see Appendix A of \cite{Alberte:2019xfh}).

\subsection{Positivity on Line theory}

\begin{figure}
\centering
\includegraphics[width=0.5\textwidth]{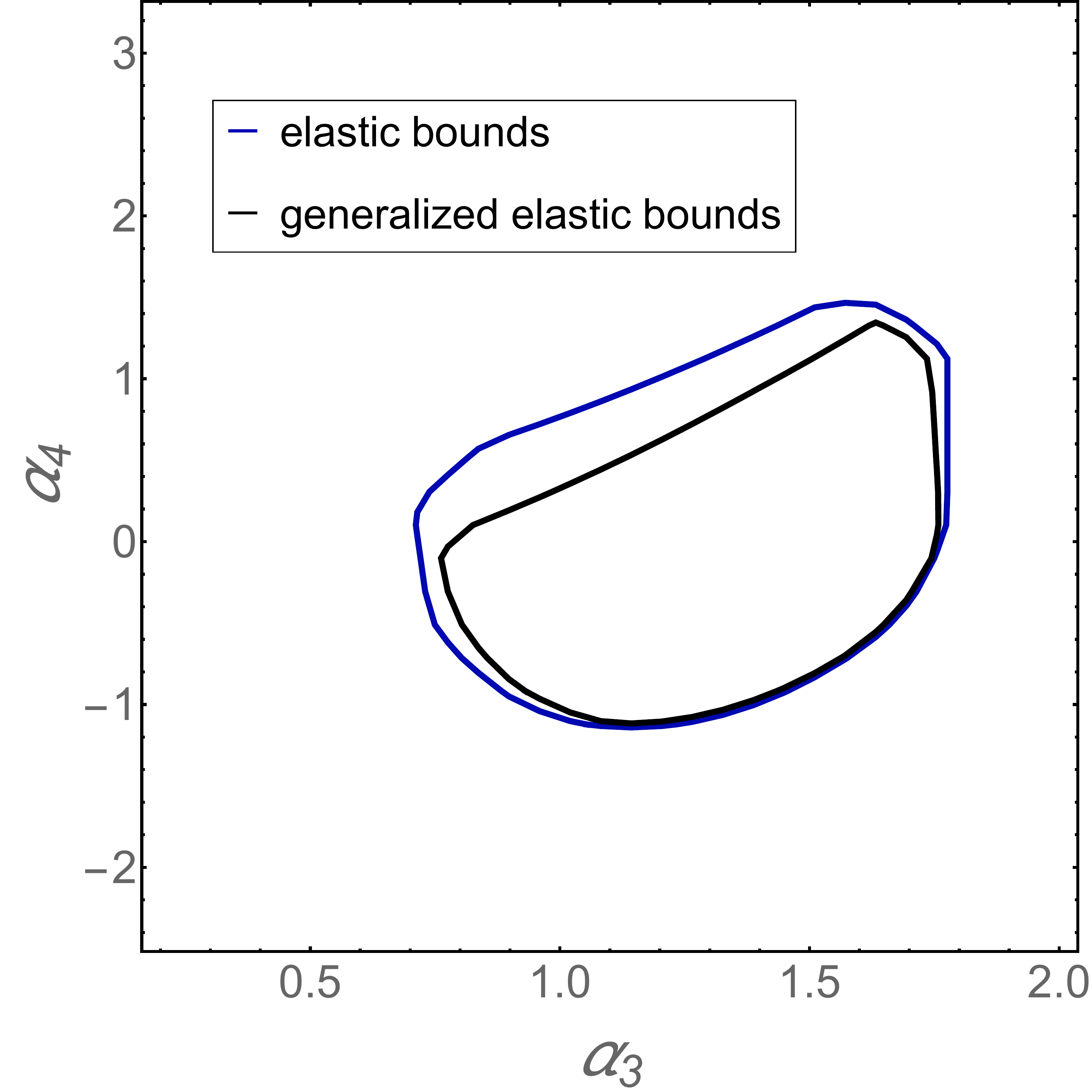}
\caption{(Line theory) Positive regions in the ($\ai_3,\ai_4$)-plane. The region within the solid blue line is allowed by the usual elastic bounds from scattering $hh\to hh$, $ff\to ff$ and $hf\to hf$, while the black line is obtained by using the generalized elastic bounds. Here we choose $\tilde{x}= \tilde m_1/\tilde m_2=2.2$, $\gi= M_1/M_2=1$ (or equivalently $\thi=0.1$), $\bi_3=1$ and $\bi_4=0$.}
\label{fig:a34}
\end{figure}

\begin{figure}
\centering
\includegraphics[width=0.46\textwidth]{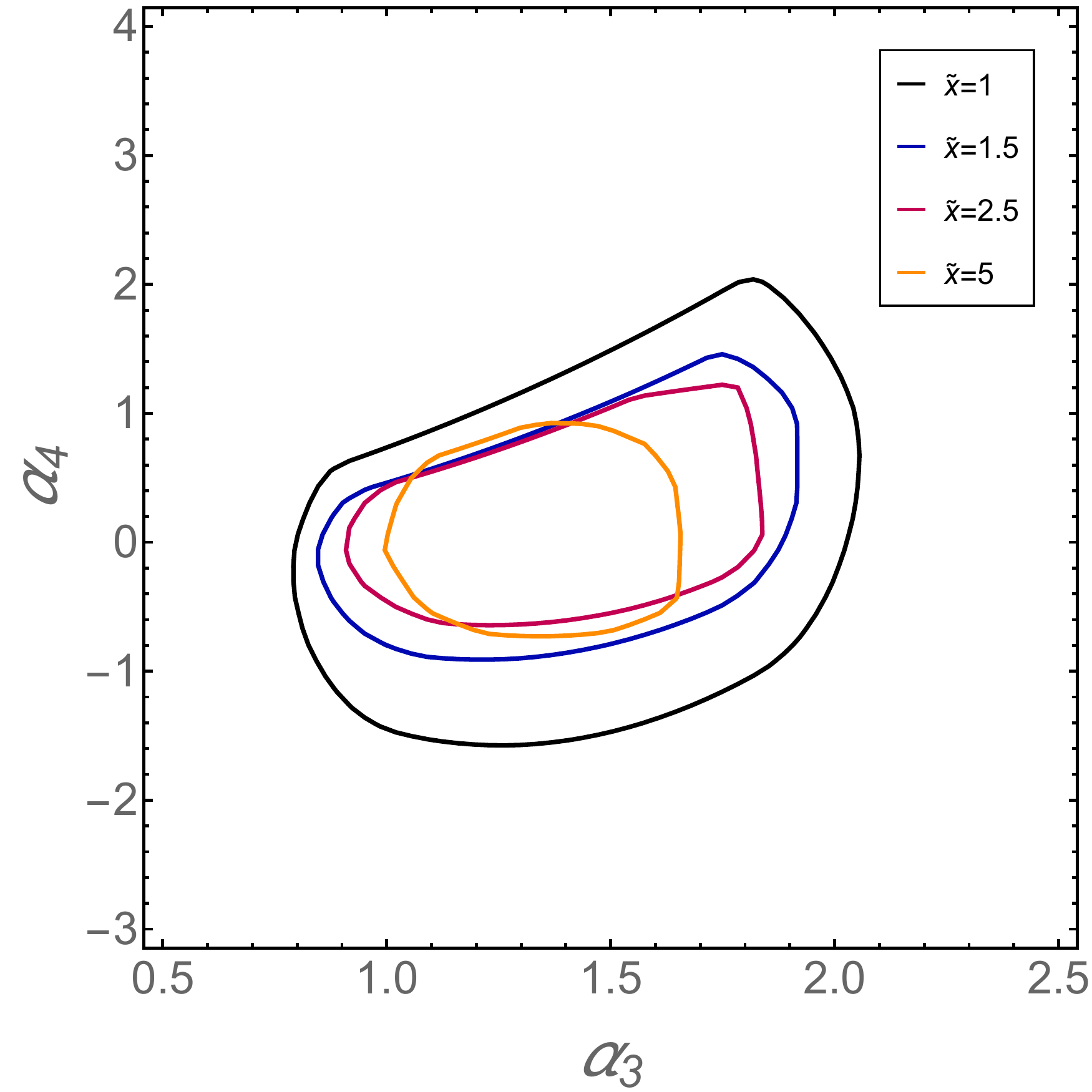}
\includegraphics[width=0.46\textwidth]{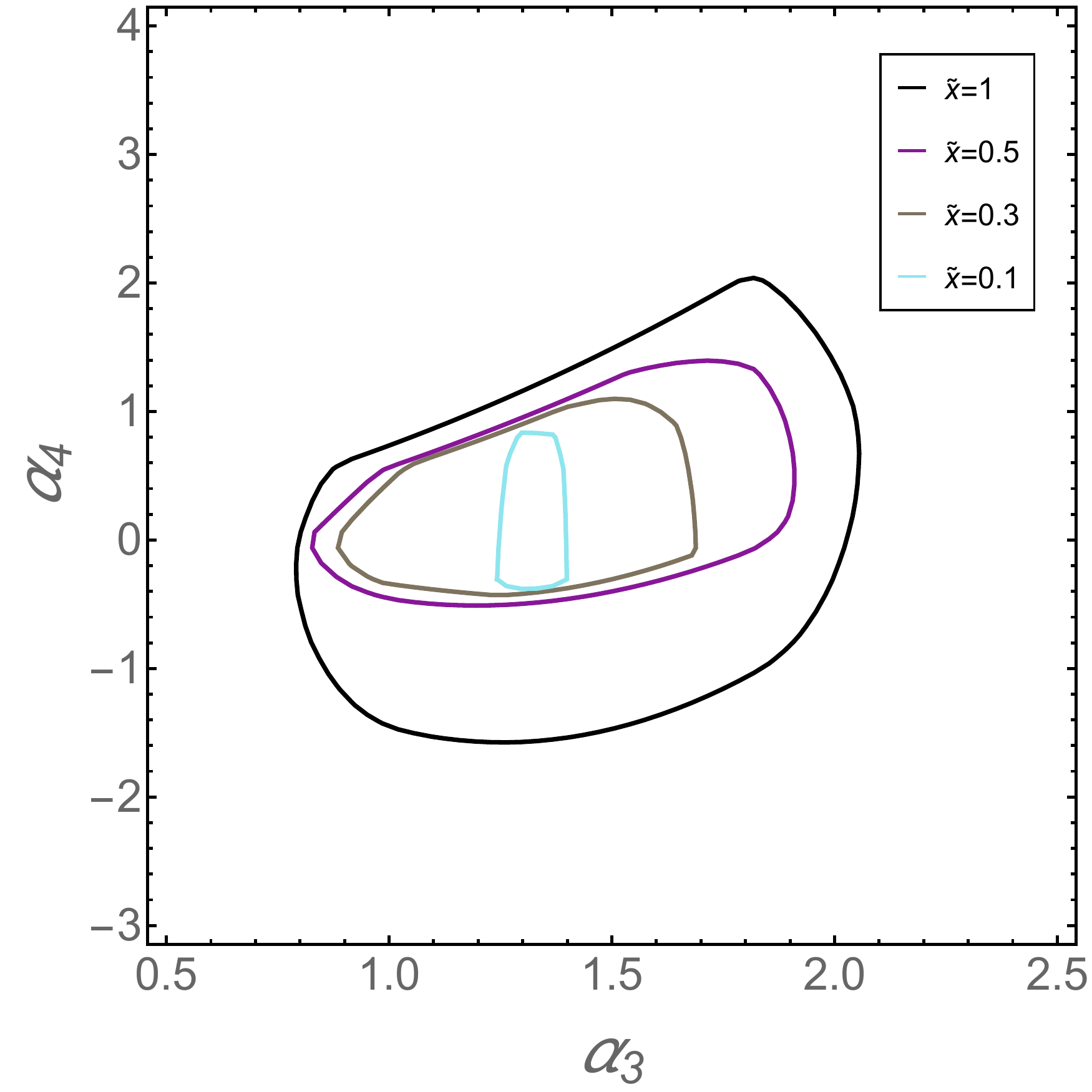}
\caption{(Line theory) Positive regions in the ($\ai_3,\ai_4$)-plane for different $\tilde x$. The regions within the solid lines are allowed by the generalized elastic positivity bounds. We choose $\bi_3=0.8$, $\bi_4=-0.1$ and $\gi=1$.}
\label{fig:ax}
\end{figure}

\begin{figure}
\centering
\includegraphics[width=0.46\textwidth]{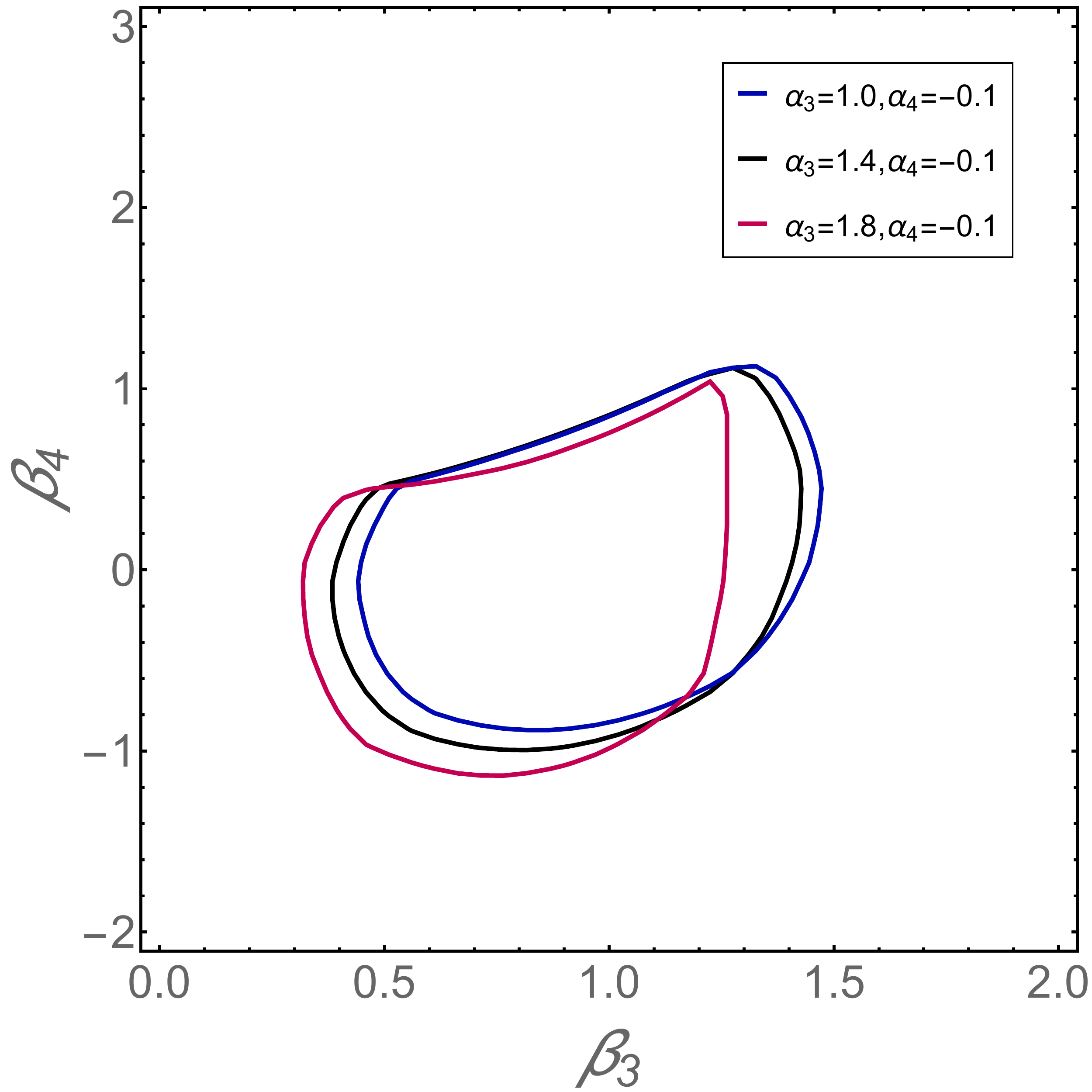}
\includegraphics[width=0.46\textwidth]{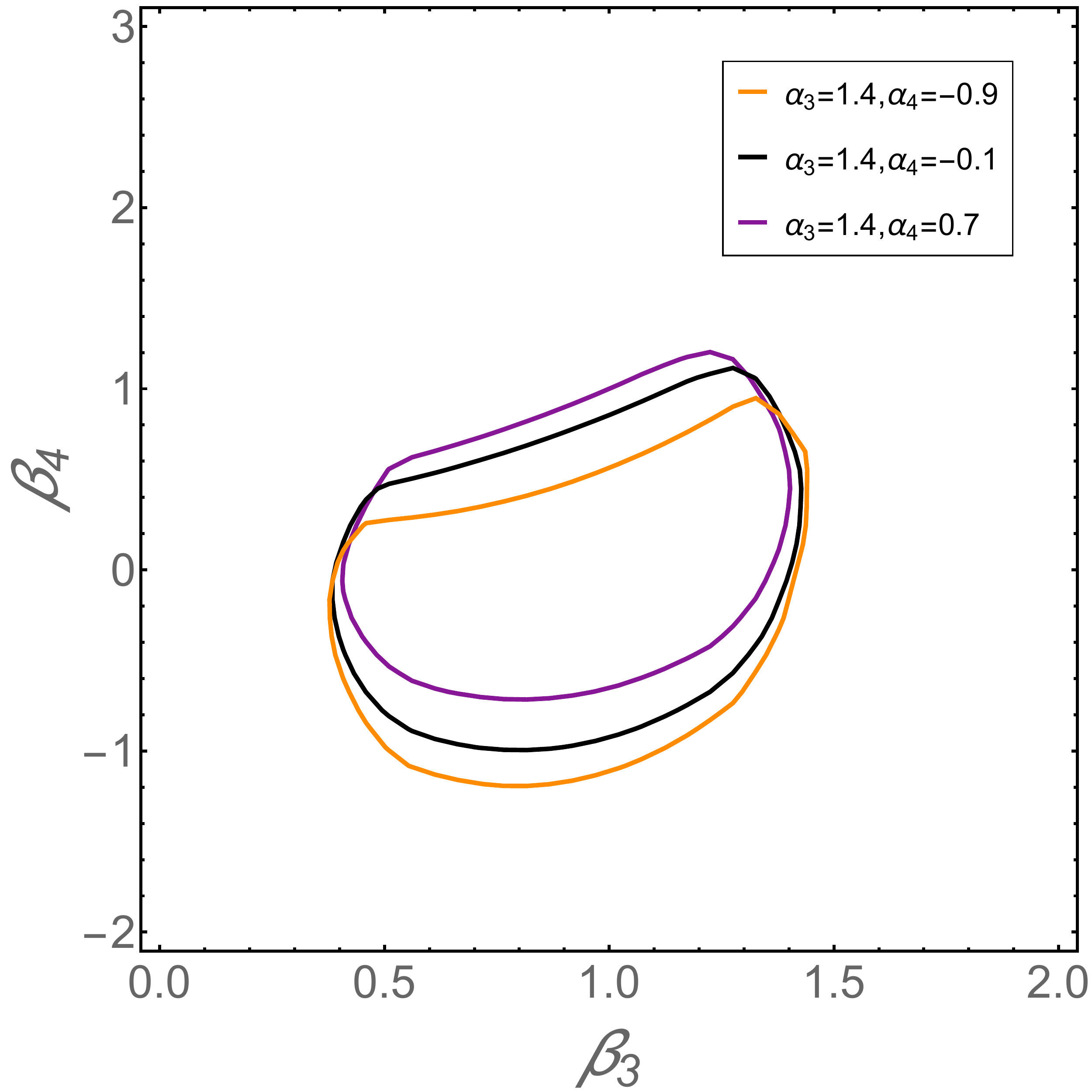}
\caption{(Line theory) Positive regions in the ($\bi_3,\bi_4$)-plane (the regions within the solid lines are allowed by the generalized elastic positivity bounds). We choose $\tilde{x}=1$ and $\gi=1$ (or equivalently $\thi=0.4$).}
\label{fig:ba}
\end{figure}

Similarly, the line theory is formally invariant under the following transformations
\bal
&h\leftrightarrow f,~\thi\leftrightarrow -\thi,~m_1\leftrightarrow m_2,~M_1\leftrightarrow M_2,\nn
&\tilde{m}_2^2 \ki_0^{(3)}\leftrightarrow\tilde{m}_2^2 \ki_3^{(3)},~
\tilde{m}_2^2 \ki_1^{(3)}\leftrightarrow\tilde{m}_2^2 \ki_2^{(3)},~
\tilde{m}_2^2 \ki_0^{(4)}\leftrightarrow\tilde{m}_2^2 \ki_4^{(4)},~
\tilde{m}_2^2 \ki_1^{(4)}\leftrightarrow\tilde{m}_2^2 \ki_3^{(4)}  .
\eal
Note that this invariance is valid only before we replace $\ki_n^{(3,4)}$ with $\tilde{\ai}_{3,4}$ and $\tilde{\bi}_{3,4}$. We can still use this property to calculate only half of the $16$ amplitudes and infer the other half from them, as in the cycle theory. Again, we need an overall factor to regularize the kinematic branch points from inelastic scattering amplitudes:
\beq
\Gi(s)=s^2\(s-4m_1^2\)^2\(s-4m_2^2\)^2\[s-(m_1+m_2)^2\]^4\[s-(m_1-m_2)^2\]^4   .
\eeq
In the line theory, the vertices in Lagrangian \eref{line} are much more complicated, so, for simplicity, we will neglect mixings between some modes and restrict to the following bounds
\bal
&\left.f^1_{\ai\bi}=f_{\ai\bi}\right|_{\ai_6\rightarrow0,\ai_7\rightarrow0,\ai_8\rightarrow0,\ai_9\rightarrow0,\ai_{10}\rightarrow0}>0  ,\nn
&\left.f^2_{\ai\bi}=f_{\ai\bi}\right|_{\ai_1\rightarrow0,\ai_2\rightarrow0,\ai_3\rightarrow0,\ai_4\rightarrow0,\ai_5\rightarrow0}>0  .
\eal
Similar to the previous cases, generalized elastic bounds can also give further constraints on the parameter space of the Wilson coefficients in the line theory. As shown in Fig.~\ref{fig:a34}, the region within the solid blue line is allowed by the usual elastic positivity bounds from scattering $hh\to hh$, $ff\to ff$ and $hf\to hf$, while, in contrast, the generalized elastic positivity bounds give rise to further constraints, which is the region inside the black line. The shrinking becomes more prominent if we allow $\tilde x= \tilde m_1/\tilde m_2$ to be different than 1, as we can see in Fig.~\ref{fig:ax}. When $\tilde{x}$ is greater (smaller) than $1$, the region allowed by positivity shrinks when $\tilde{x}$ decreases (increases). The limit of $\tilde{x}\rightarrow 0$ (with $\gi$ fixed) corresponds to when $f_{\mu\nu}$ becomes almost massless and $h_{\mu\nu}$ remains massive. We find that in such limit the region shrinks to a point.  Fig.~\ref{fig:ba} shows that how the region allowed by the generalized elastic positivity in the ($\bi_3,\bi_4$)-plane varies with different $\ai_3$ and $\ai_4$. As we see that changing these parameters mostly amounts to slightly shifting the positive region in the ($\bi_3,\bi_4$)-plane. So the positivity constraints on the dRGT potential that mixes $h_{\mu\nu}$ and $f_{\mu\nu}$ (the $\bi_{3}$ and $\bi_{4}$ terms) are insensitive to the dRGT potential of $h_{\mu\nu}$ (the $\ai_{3}$ and $\ai_{4}$ terms).

\section{Generalizations to multiple fields}

\label{sec:multifield}

In the previous sections, for simplicity and also for the clarity of arguments, we have restricted ourselves to cases where there are only two dynamical spin-2 fields (or three metrics if the Minkowski metric is also counted as a background field). Many of the results obtained analogously apply to cases with multiple fields. While we do not wish to repeat the same calculations for the multiple field cases in this section, we will point out that generalized elastic positivity bounds can give rise to new constraints in these generalizations, in comparison to the usual elastic positivity bounds.

\subsection{Pseudo-linear theory}

Let us first consider the pseudo-linear theory, whose leading terms have been completely excluded by generalized elastic positivity for the bi-field case. In the pseudo-linear theory with more than two fields, there will be new kinds of terms, which involve, say, three or four kinds of fields in a vertex. We will see that in this case, the generalized elastic positivity bounds can also provide extra constraints on these new terms.

\begin{figure}
\centering
\includegraphics[width=0.28\textwidth]{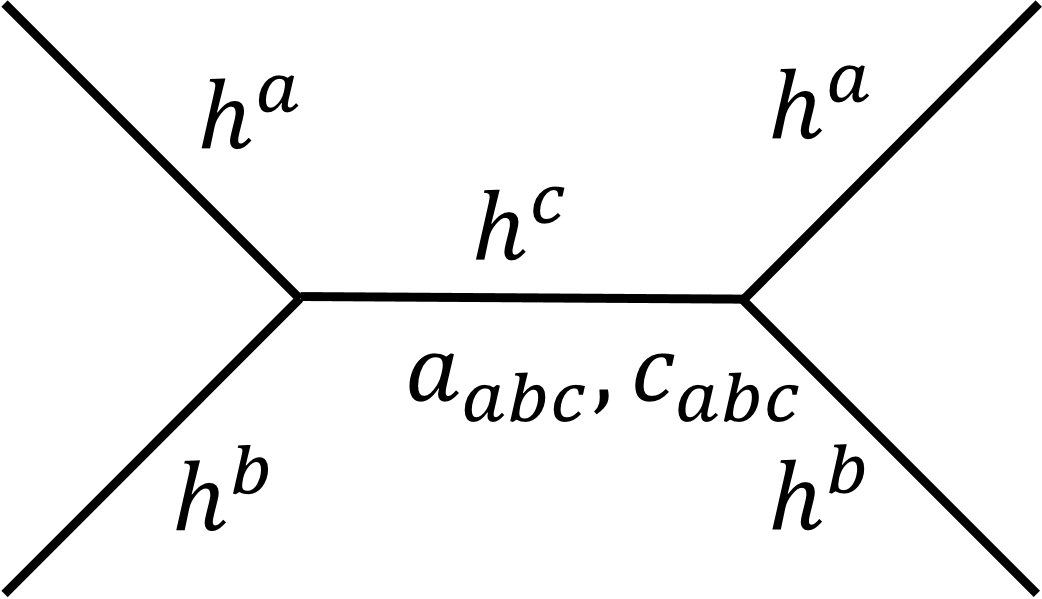}
~~~~~~~~~~~~~~~~~
\includegraphics[width=0.16\textwidth]{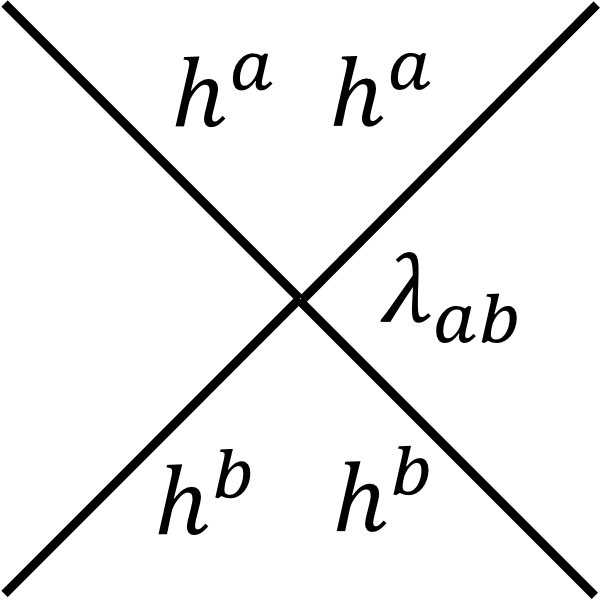}
\caption{Feynman diagrams for elastic scattering $h^a h^b \to h^a h^b$ ($a\neq b$) in the pesudo-linear theory. The left diagram contributes $a_{abc}$ and $c_{a,b,c}$ to the positivity bound and the right one contributes $\li_{ab}$ to the bound.}
\label{fig:PLF}
\end{figure}

Suppose we have a number of spin-2 fields $h^a=h^a_{\mu\nu}$ with mass $m_a$, $a=1,2,...,N$. Analogous to the bi-field case, from elastic scattering $h^a h^a\rightarrow h^a h^a$, positivity excludes the following types of vertices: $\epi\epi I h^a h^a h^a$, $\epi\epi h^a h^a h^a h^a$, $\epi\epi I h^a h^a h^b$, $\epi\epi\pd^2 h^a h^a h^a$, $\epi\epi\pd^2 h^a h^a h^b$ and $\epi\epi\pd^2 h^b h^a h^a$, where $a\neq b$. So for the highest cutoff theory, we are left with Lagrangian
\bal
g_*^2 \mc{L}^{\rm multi}_{\rm pseudo} = &\sum_{a}\mc{L}_{\mathrm{FP}}(h^a,m_a)+\frac{1}{M}\(\sum_{a,b,c}a_{abc}\epi\epi\pd^2 h^a h^b h^c+\sum_{a,b,c}m^2 c_{abc}\epi\epi I h^a h^b h^c\)\\
+\frac{m^2}{M^2} & \(\sum_{a,b}\li_{ab}\epi\epi h^a h^a h^b h^b
+\sum_{a,b}d_{ab}\epi\epi h^a h^a h^a h^b+\sum_{a,b,c}p_{abc}\epi\epi h^a h^a h^b h^c +\sum_{a,b,c,d}q_{abcd}\epi\epi h^a h^b h^c h^d\) ,
\eal
where $a$, $b$, $c$ and $d$ are all different and $a_{abc}$, $c_{abc}$, $\li_{ab}$, $d_{ab}$, $p_{abc}$, $q_{abcd}$ are Wilson coefficients. All of these vertices contribute to some tree level scattering amplitudes. As visualized in Fig.~\ref{fig:PLF}, elastic positivity bounds from $h^a h^b \to h^a h^b$ can give rise to constraints on $a_{abc}$, $c_{abc}$ and $\li_{a,b}$. The $\li_{a,b}$ vertex is excluded entirely in the bi-field case, but now it re-emerges in the case with more fields because of the mixing with the $a_{abc}$, $c_{abc}$ terms in the $h^a h^b \to h^a h^b$ amplitude, thus not excluded with the previous argument. If we only consider the usual elastic positivity bounds, apart from those vertices already excluded by $h^a h^a\rightarrow h^a h^a$, we can now constrain the $a_{abc}$, $c_{abc}$ and $\li_{a,b}$ vertices. On the other hand, if we use the generalized elastic positivity bounds, we can additionally constrain all the remaining Wilson coefficients $d_{ab}$, $p_{abc}$ and $q_{abcd}$, as the generalized elastic positivity bounds make use of elastic and inelastic scattering amplitudes. Whether generalized positivity bounds can exclude general multi-field pseudo-linear theory in 4D and beyond is left for future work.

\subsection{$\Lambda_3$ cycle theory}

Now, let us turn to the $\Lambda_3$ cycle theory.  As discussed in Section \ref{sec:Li3theory}, the generalized elastic positivity bounds can not give further restrictions on the Wilson coefficients in the case of bi-field $\Lambda_3$ cycle theory. The reason for this is because in that case under the $\Li_{7/2}$ to $\Li_{3}$ re-scaling all information from inelastic scattering gets scaled to be subleading. Here we will show that, for cases with more than two fields, generalized elastic positivity bounds do give rise to new constraints.

\begin{figure}
\centering
\includegraphics[width=0.16\textwidth]{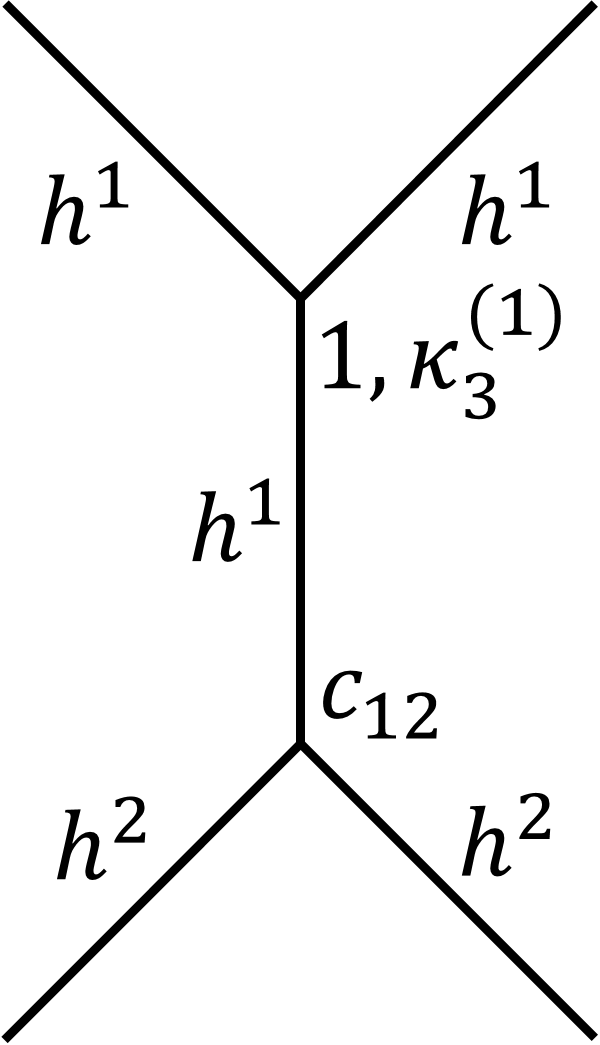}
~~~~~~~~~~~~~~
\includegraphics[width=0.16\textwidth]{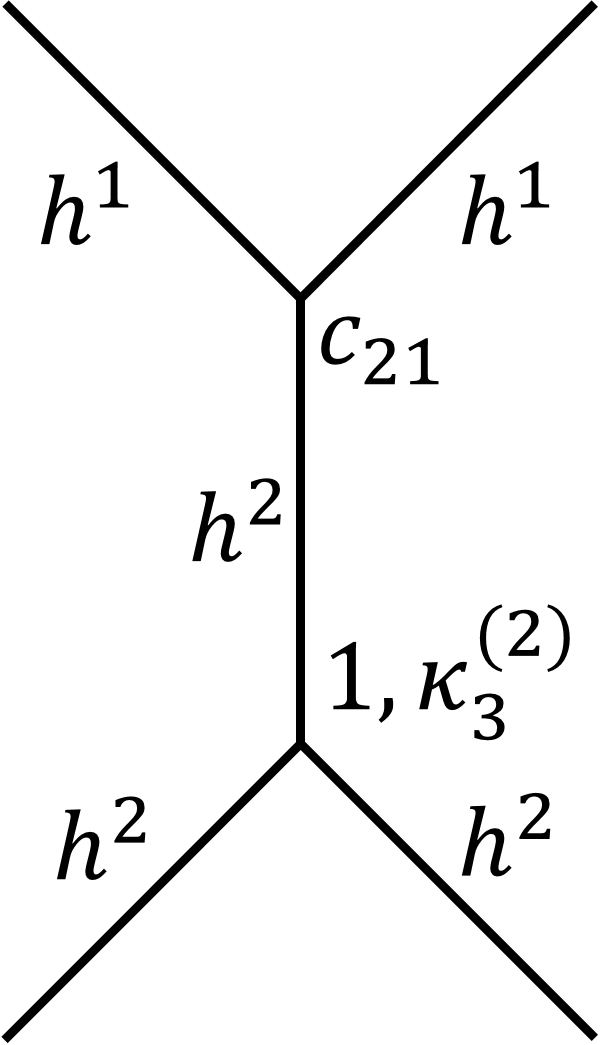}
~~~~~~~~~~~~~~
\includegraphics[width=0.16\textwidth]{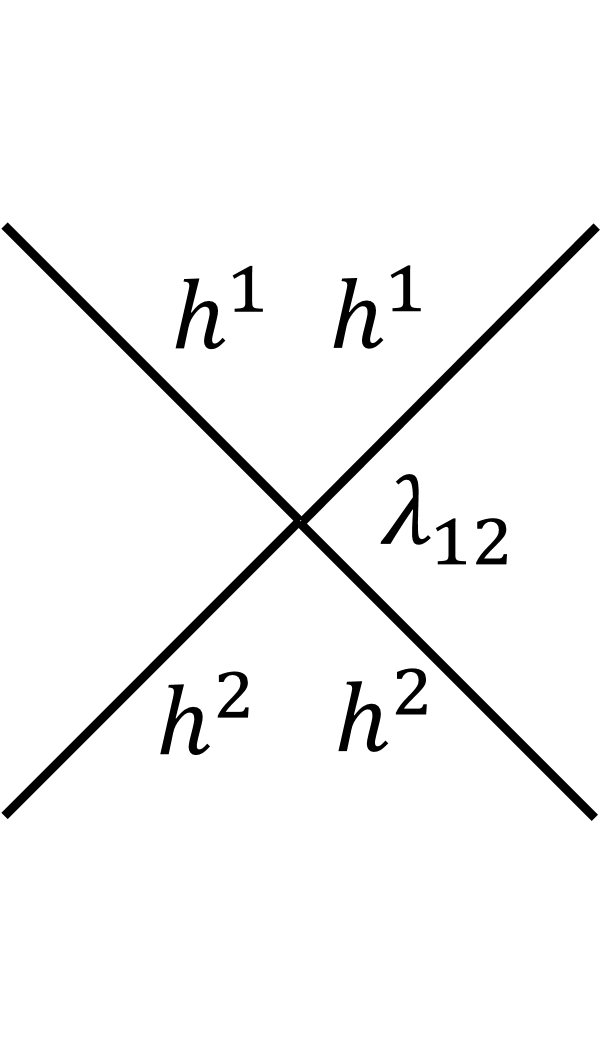}
\caption{Leading Feynman diagrams for elastic scattering $h^1 h^2 \to h^1 h^2$ in a bi-field or multi-field $\Li_3$ cycle theory.}
\label{fig:L3F1}
\end{figure}

The interacting Lagrangian for a multi-field cycle theory includes the following terms
\bal
&g_*^2\mc{L}_{\rm cycle}^{\rm multi} \supset\frac{m^2}{M}\(\sum_{a,b} c_{ab}\epi\epi I h^a h^b h^b+\sum_{a,b,c} t_{abc}\epi\epi I h^a h^b h^c\)\nn
+&\frac{m^2}{M^2}\(\sum_{a,b}\li_{ab}\epi\epi h^a h^a h^b h^b
+\sum_{a,b}d_{ab}\epi\epi h^a h^a h^a h^b+\sum_{a,b,c}p_{abc}\epi\epi h^a h^a h^b h^c+\sum_{a,b,c,d}q_{abcd}\epi\epi h^a h^b h^c h^d\)  ,
\eal
where $a$, $b$, $c$ and $d$ are all different. Again, by choosing the interactions between different fields to be suppressed by an extra factor of $m/\Lambda_3$, we get a $\Li_3$ cycle theory. This means that $c_{ab}$, $t_{abc}$, $\li_{ab}$, $d_{ab}$, $p_{abc}$ and $q_{abcd}$ is much smaller than $\mc{O}(\ki_{3,4}^{(a)})\sim 1$. Thus, to effectively constrain these coefficients, we need to have positivity bounds that start at $\mc{O}(m/\Lambda_3)$ not at $\mc{O}(1)$. Let us consider a $\Lambda_3$ cycle theory with $3$ fields $h^1$, $h^2$ and $h^3$ for simplicity. The usual elastic bounds from the $h^1 h^2 \to h^1 h^2$ scattering (the leading Feynmann diagrams shown in Fig.\ref{fig:L3F1}) can give rise to constraints on $c_{12}$, $c_{21}$ and $\li_{12}$, and similarly the $h^1 h^3 \to h^1 h^3$ scattering can give rise to constraints on $c_{13}$, $c_{31}$ and $\li_{13}$. On the other hand, if we consider generalized elastic positivity bounds from the $h^1 X \to h^1 X$ scattering where $X$ is a superposition of $h^2$ and $h^3$ without $h^1$, the positivity bound will contain information about $h^1 h^2 \to h^1 h^2$, $h^1 h^3 \to h^1 h^3$ and $h^1 h^2 \to h^1 h^3$, which are all at the $\mc{O}(m/M)$ and give rise to extra constraints on $t_{123}$ and $p_{123}$. (In comparison, if we only have two fields, generalized positivity bounds from the $h^1 X \to h^1 X$ scattering where $X$ is a superposition of $h^1$ and $h^2$ will have to contain $h^1 h^1 \to h^1 h^1$, whose leading order is at $\mc{O}(1)$, while the contributions from $h^1 h^2 \to h^1 h^2$ and $h^1 h^1 \to h^1 h^2$ start at $\mc{O}(m/\Lambda_3)$.)

\begin{figure}
\centering
\includegraphics[width=0.16\textwidth]{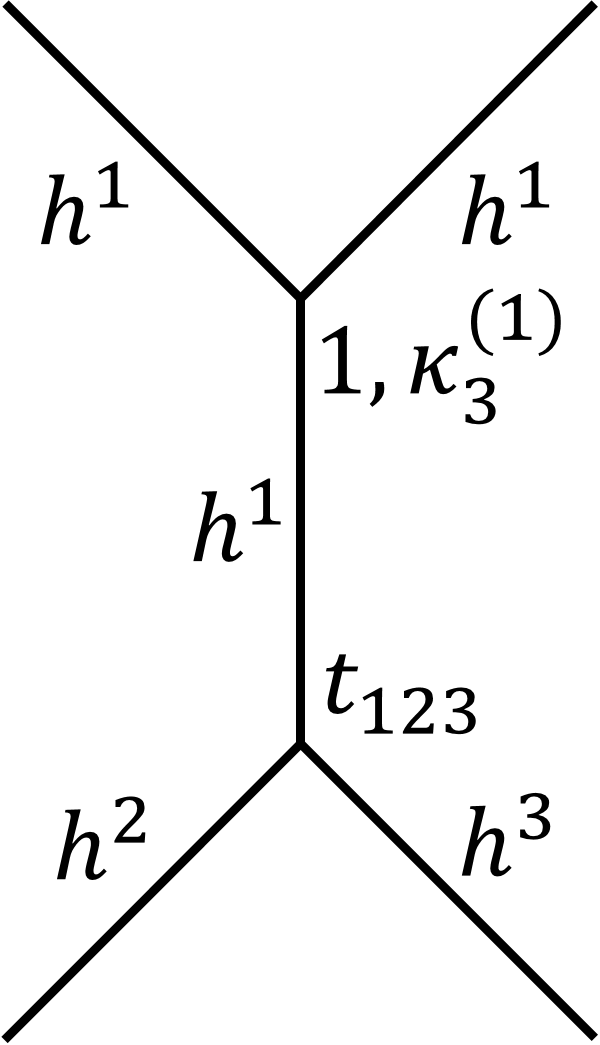}
~~~~~~~~~~~~~~~~~
\includegraphics[width=0.16\textwidth]{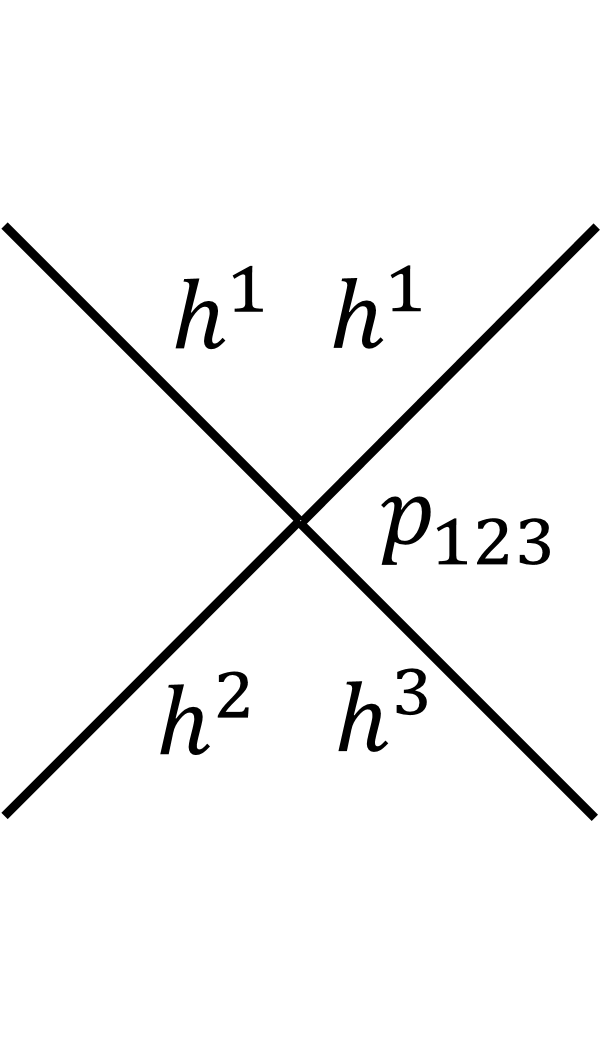}
\caption{Leading Feynman diagrams for inelastic scattering $h^1 h^2 \to h^1 h^3$ in a multi-field $\Li_3$ cycle theory.}
\label{fig:L3F2}
\end{figure}

\acknowledgments

We would like to thank Zong-Zhe Du, Andrew Tolley and Yang Zhang for helpful discussions. 
 CZ is supported by IHEP under Contract No.~Y7515540U1, and by
National Natural Science Foundation of China (NSFC) under grant No.~12035008.
SYZ acknowledges support from the starting grants from University of Science
and Technology of China under grant No.~KY2030000089 and GG2030040375, and is
also supported by NSFC under grant No.~11947301 and 12075233.

\bibliographystyle{JHEP}
\bibliography{refs1}

\end{document}